\documentclass[aps,prb,11pt,longbibliography]{revtex4-2}

\usepackage{amsmath,amssymb,mathrsfs}
\usepackage{graphicx}   %
\usepackage{verbatim}   %
\usepackage{color}      %
\newcommand{\QE}{{\sc Quantum Espresso}}
\newcommand{\ATS}{{\sc ATSUP}}
\begin{document}

\title{Challenges in predicting positron annihilation lifetimes in lead halide perovskites: correlation functionals and polymorphism}
\author{Kajal Madaan}
\affiliation{Université Paris-Saclay, CEA, Service de Recherches en Corrosion et Comportement des Matériaux, SRMP, 91191 Gif sur Yvette, France}
\author{Guido Roma}
\affiliation{Université Paris-Saclay, CEA, Service de Recherches en Corrosion et Comportement des Matériaux, SRMP, 91191 Gif sur Yvette, France}
\email{guido.roma@cea.fr}
\author{Jasurbek Gulomov}
\affiliation{Université Paris-Saclay, CEA, Service de Recherches en Corrosion et Comportement des Matériaux, SRMP, 91191 Gif sur Yvette, France}
\author{Pascal Pochet}
\affiliation{Department of Physics, IriG, Univ. Grenoble-Alpes and CEA, Grenoble,France}
\author{Catherine Corbel}
\affiliation{CEA/DRF/IRAMIS, University Paris-Saclay, Gif-sur-Yvette, France}
\author{Ilja Makkonen}
\affiliation{Department of Physics, University of Helsinki, P.O. Box 43, FI-00014 Unicversity of Helsinki, Helsinki, Finland}

\begin{abstract}
  Halide perovskites have emerged in the last decade as a new important class of semiconductors for a variety of optoelectronic applications.
  A lot of previous studies were thus devoted to the characterisation of their point defects.
  Positron annihilation spectroscopy is a well recognized tool for probing vacancies in materials. Recent applications
  of this technique to APbX$_3$ halide perovskites are sparse, and the rare theoretical predictions of positron lifetimes
 in these materials, published in association with experiments, do not fully agree with each other. These works suggest that
 vacancies on the A site are not detected.

 In our theoretical study we focus on the role of the electron-positron correlation functional.
 We thoroughly revisit and compare several approximations
 when applied to methylammonium lead iodide (MAPbI$_3$) with or without vacancies,
 as well as inorganic perovskites CsPbI$_3$ and CsPbBr$_3$, in various phases.
 We show also the relationship between the size of the voids, through Voronoi volumes, and the calculated lifetimes.
 For the cubic phases we investigate in detail the role of polymorphism, including the distribution of vacancy formation energies and positron annihilation lifetimes.
 
In our lifetimes calculations, apart from older and more recent semi-local approximations for the electron-positron correlation potential, we also consider the weighted density approximation (WDA), which is truly non-local and should better describe positron annihilation in regions with strong electronic density variations.
We show that for this class of materials, and especially for cations vacancies, the influence of the chosen approximation is crucial, much stronger than in metals, alloys and conventional semiconductors. This influence may induce to reconsider the interpretation of experimentally determined lifetimes.

\end{abstract}

\maketitle

\section{Introduction}

Hybrid and fully inorganic halide perovskite (HP) are now on the verge of the market for solar cells and other applications.\cite{perovskite-info} Besides solar cells, which first sparked, and still fuel, a huge interest in this class of materials, many other applications, related mainly to their optoelectronic properties, can benefit from their surprising physical properties.\cite{Dong_2023}  
Being able to characterize point defects and their concentration in halide perovskites is crucial to check the quality of absorbing layers of photovoltaic devices and to advance in understanding the basic mechanisms of carrier recombination and phase stability. A variety of optical methods have been very frequently employed\cite{Srivastava_2023}, together with other more or less indirect techniques.

Positron annihilation lifetime spectroscopy is a well established tool to probe vacancy defects, with a fairly high sensitivity with respect to other spectroscopic techniques. It has commonly been used to detect such defects and determine their concentrations in metals\cite{Eldrup_2003,Selim_2021} and semiconductors.\cite{MakkonenTuomisto_2024}

A large number of papers has been devoted to first principles calculations of various defect properties in halide perovskites~\cite{Guo_2024}. The few theoretical predictions of positron lifetimes in these material are published in papers reporting also experimental determinations of positron lifetimes.\cite{Keeble_2021,Musiienko_2022,Ni_2024,Schmidt_2025}
The reason is that it is  difficult to find experimental evidence on which to rely for linking the experimental positron lifetimes to specific positron annihilation quantum states in HPs, i.e., delocalised in the lattice versus localised in point defects or other microstructural features.
The authors consequently established such link by comparing experimental and calculated lifetimes.  

This paper focusses on the choice of the approximation used in the positron lifetime calculations for the electron-positron correlation functional (EPCF). It shows that this choice has a strong effect on the positron lifetime values for positron delocalised in the HP crystalline lattice as well as localised at primary intrinsic vacancies. The mini-review presented below deals with the reported experimental lifetimes in HPs in a first part and with the calculations in a second one.  The second part shows that the key role of the EPCF for the description of the positron screening cloud has been previously overlooked. In the discussion, the comparison between our calculated values and the experimental values quoted in the mini-review leads to the conclusion that the range of  experimental and EPCF dependent values in HPs is too wide to establish an optimal choice of EPCF.

Recent experimental works have dealt with positron lifetimes in halide perovskites.\cite{Dhar_JPCC2017,Dhar_JPCL2017,Dhar_2018,Keeble_2021,Musiienko_2022,Moshat_2023,Ni_2024,Schmidt_2025,Cai_2025,CCa,CCb} 
Depending on whether a $\beta^+$ radioactive ${}^{22}$Na
source or a slow positron beam is used, a few papers report positron
lifetimes measured either deep within the bulk\cite{Dhar_JPCC2017,Dhar_JPCL2017,Dhar_2018,Musiienko_2022,Moshat_2023,Schmidt_2025},  near the surface of halide perovskites,\cite{Keeble_2021,Ni_2024} or at varying depth.\cite{Cai_2025,CCa,CCb}
This is probably one of the reason of the wide spread of the data, together with the characteristics of the samples.
We provide a detailed analysis of these works in the Supplemental Material\cite{SuppInfo}  (section A and Fig. S1).

The positron implantation mean depths for MAPbI$_3$ are $\sim$55 $\mu$m for ${}^{22}$Na\cite{Dryzek_2006} and $\sim$1.66 $\mu$m for 25~keV positron beam, which is the maximum energy 
of most of the slow positron beams in operation\cite{Krause_1999}.
A general remark about the positron decay lifetimes reported in literature is that none of the
experimental investigations lead the authors to report positron decay lifetimes arising from the halide perovskite itself ---MAPbI$_3$, MAPbBr$_3$ and CsPbBr$_3$--- longer than 500~ps. This gives evidence that the free positron annihilation mode
where positrons are in interaction with the electronic cloud\cite{Krause_1999} is the dominant
annihilation mode in halide perovskites. Halide perovskites are semi-conductors where the electronic density is high enough to prevent the positronium (Ps) annihilation mode where positron-electron pairs are in interaction with the electronic cloud. We will provide further evidence of this point later.

Some theoretical predictions of positron lifetimes in halide perovskites, all of which based on two components density functional theory, have been published in recent years.
Three of them use the approach implemented in the ABINIT software package\cite{ABINIT2020,Wiktor_2015}, with a self-consistent scheme including the positron and electron densities as well as the relaxation of the atomic positions. The fourth, by Musiienko {\sl et al.},\cite{Musiienko_2022} devoted to bromides, uses the conventional scheme, where the response of the electron density to the presence of the positron is implemented via a function of position called the enhancement factor.\cite{PuskaNieminen_RMP1994}

Keeble {\sl et al.}\cite{Keeble_2021} show that the calculated positron lifetimes values depend on the approximations used to determine the electronic density, and on the self-consistency of the positron density with the former and even with the atomic relaxations. For positrons delocalised in the
MAPbI$_3$ pristine lattice, the calculated lifetimes vary from 342 to 353~ps. For localised positrons, the calculated values vary from 360 to 369~ps for the negative lead vacancy, V$_{Pb}^{-2}$, and, over a much wider range, from 401 to 442~ps for the methylammonium vacancy, V$_{MA}^-$. 

Within a computational scheme similar to Keeble {\sl et al.}, Schmidt {\sl et al.} report results for MAPbI$_3$ which differ by, at most, 10~ps (see Ref.~\cite{Schmidt_2025}, Supporting Informations).

For MAPbBr$_3$, Musiienko {\sl et al.} predict that lifetimes increase from 333.4~ps for positrons delocalised in lattice to 342.9, 347.0 and 357.9~ps for positrons localised at Br, Pb and MA vacancies, respectively.
The vacancies are, in this case, in the neutral state. 
For the same material Ni {\sl et al.} obtain lifetimes of 312~ps for the positron delocalised in the lattice and  348~ps and 440~ps for positron localised in lead and methylammonium vacancies, with charges -2 and -1, respectively.\cite{Ni_2024}. The lifetimes obtained by Schmidt {\sl et al.} for MAPbBr$_3$, using the same approach differ by less than 2\%.
The difference between the latter two papers and the one by Musiienko are more significant and may depend, for the delocalised positron, from the computational scheme, and for the localised ones also from the charge state of the vacancies. 

In CsPbBr$_3$, the positron lifetimes calculated by Musiienko {\sl et al.}\cite{Musiienko_2022} 
 with the same approach as
for MAPbBr$_3$, yields 355.1~ps for the pristine lattice while increasing to 364.8, 371.9 and 394.8~ps respectively 
in Cs, Pb, and Br neutral vacancies.

The comparison between the lifetimes by Musiienko {\sl et al.}\cite{Musiienko_2022} for MAPbBr$_3$
and CsPbBr$_3$ shows that the replacement of the Cs inorganic cation by the organic MA cation has a huge effect on
the overlap between the electronic and positron density. The vacancy associated to the longest lifetime switches
from the MA vacancy in MAPbBr$_3$ to the Br vacancy in CsPbBr$_3$.
The comparison between the experimental and computed positron lifetimes show that the longer decay lifetimes
resolved for near-surface positron lifetime spectra are all longer than the computed positron lifetimes in the pristine lattice
for the compounds MAPbI$_3$ and MAPbBr$_3$. For the spectra recorded deep within the bulk, Dhar {\sl et al.}\cite{Dhar_JPCC2017,Dhar_JPCL2017} reports
values of 309$\pm$5, 326$\pm$3, 335$\pm$3)~ps for the longer decays, below the computed lattice lifetimes in MAPbI$_3$. Schmidt
{\sl et al.}\cite{Schmidt_2025} report values, 388 ps, above the computed positron annihilation lifetime in the MAPbBr$_3$ lattice. The wide
ranges of experimental decay lifetimes show that positrons are strongly captured in open-volume defects in
the halide perosvkites so far characterised using positron annihilation lifetime spectroscopy.

Previous works on positron annihilation in halide perovskites did not specifically focus on the theoretical side.
Keeble {\sl et al.} show some interesting comparison between different calculation schemes ---involving the influence of the electron density and the atomic relaxations. However, the crucial choice of the electron-positron correlation functional (EPCF), that enhances the electronic density that overlaps with the positron density, remains to be investigated.

Early calculations of positron lifetimes and electron momentum distributions in elemental metals,\cite{Gupta_1980,Puska_1983} Si and III-V semiconductors\cite{PuskaCorbel_1988}, addressed the issue of the EPCF. In those calculations, the authors generally used semilocal correlation functionals. After the Boronski-Nieminen (BN) parametrization\cite{BoronskiNieminen_1986} based on the LDA, which has been used in many studies since its appearance, the question of an appropriate EPCF to be used with gradient corrected DFT calculations was addressed in the 90's\cite{Barbiellini_1995}. Various improvements lead to a parameter free version\cite{Barbiellini_2015} which has been applied to various materials since.
Various approximations including the latter have recently been compared to predictions of lifetimes based on a Quantum Monte Carlo many body wave function describing both electrons and the positron.\cite{Simula_2022} A different approach to go beyond semilocal correlation functionals was chosen to investigate positron annihilation at CdTe quantum dots surfaces,\cite{Shi_PositronCdSeWDA_PRL2018} by using a truly non-local functional, a recently revisited version\cite{Callewaert_2017} of the classical weighted density approximation (WDA). This approach is expected to better represent systems with significant inhomogeneities like surfaces and voids, which represent a challenge for the prediction capabilites of positron lifetimes calculations.

Here we address the issue of the choice of the EPCF for the hybdrid perovskite MAPbI$_3$ in various phases, including vacancies, especially in the tetragonal phase, which is the stable one at room temperature.\cite{Whitfield2016} We also consider as a comparison some inorganic counterparts (CsPbI$_3$ and CsPbBr$_3$). For these systems we performed lifetimes predictions using different EPCFs. We used also the non-local weighted density approximation (WDA). The range of calculated lifetimes for the same structure is in general much larger than the expected experimental error on lifetimes measurements, which warns about the risks of relying  on lifetimes predicted with a single EPCF scheme for interpreting experimental results.

Another issue we address here is the influence of polymorphism on defect properties by calculating the distribution of formation energies and positron lifetimes. To the best of our knowledge, this aspect has not been previously investigated in detail.

The paper is organized as follows: after a methodological section \ref{Methods} we present the results in the following order: pristine phases (i.e., without defects, section \ref{StablePhases}), vacancies in tetragonal MAPbI$_3$ (section \ref{Vacancies}) and we discuss the various approximations for the EPCF. In section \ref{PolymorphousCubic} we present and discuss results obtained for polymorphous supercells of the cubic phase of MAPbI$_3$.
In section \ref{Discussion} we discuss the implication of our results in connection with experimental findings.
We summarize and conclude in section \ref{Conclu}.

\section{Methods}
\label{Methods}

Our approach for the prediction of positron lifetimes is based on two-component DFT (2CDFT)~\cite{PuskaNieminen_RMP1994}. We first compute the self-consistent Kohn-Sham electronic density using the \QE\ software package;\cite{QE-2009,QE-2017,QE_GPU} the electronic density is then used in the \ATS\ (formerly MIKA-Doppler) software~\cite{Torsti_2006} to compute the lifetime in the zero-positron density limit. In this scheme the modification of the electron density in the vicinity of the positron, due to its presence, is represented through the so called enhancement factor, $\gamma(r)$, which is generally parametrized on sophisticated theoretical approaches of model electron/positron systems.  Several electron-positron correlation potentials are considered in \ATS\ giving rise to corresponding enhancement factors, used in the expression of the positron annihilation rate. We explored several approximations: two LDA based ones, by Boronski and Nieminen (BN-LDA \cite{BoronskiNieminen_1986}) and Drummond {\sl et al.} (D-LDA \cite{Drummond_2011}), three GGA based (B95-GGA by Barbiellini {\sl et al.}\cite{Barbiellini_1995}, K14-GGA by Kuriplach and Barbiellini\cite{Kuriplach_2014}, and B15-GGA by Barbiellini and Kuriplach\cite{Barbiellini_2015}), and the non-local weighted density approximation in the recent version by Callewaert {\sl et al.} \cite{Callewaert_2017}, with shell partitioning.
For B95-GGA we stick to the suggested parameter $\alpha=0.22$ and for the K14-GGA to $\alpha=0.05$ together with D-LDA parametrization. For the WDA, although for insulating materials the screening charge Q should be less than one, the presence of semicore electrons in our pseudopotentials would suggest values of Q well beyond one.\cite{Callewaert_2017} We varied the screening charge Q from Q=0.8 to Q=2.0 to investigate its effect on the calculated lifetimes. Q=2 corresponds to the limit of vanishing electron density in a homogeneous electron gas, where the formation of the negative positronium ion, Ps$^\textrm{-}$ is expected.

Our DFT calculations employed the optB86b+vdW exchange-correlation (xc) functional proposed in Ref.~\onlinecite{Klimes_2011}. We use ultrasoft pseudopotentials with 14 electrons kept in valence for Pb, 9 for Cs, 17 for Br, and standard valence for the other elements.
The technical details for the \QE\ plane waves calculations include large kinetic energy (charge density) cutoffs of 70 Ry (560 Ry) for MAPbI$_3$ and 90 Ry (720 Ry) for CsPbI$_3$ and CsPbBr$_3$. We used 6$\times$6$\times$6 $\Gamma$ centered grids of {\bf k}-points for the primitive unit cells of the three materials containing one formula unit. The {\bf k}-points grids used for larger cells are equivalent to those for the unit cells, except for the 16 and 32 MAPbI$_3$ formula units (fu) cells, for which we used 1$\times$1$\times$2 and $\Gamma$ only grids, respectively. 

Halide perovskites present the peculiarity that, beside clearly ordered phases, occurring at low temperature, higher temperature phases, although frequently described as ordered average structures, include disorder whose nature is not only due to dynamics.
The structure of the various ordered phases of the inorganic compounds were taken from our previous works on CsPbI$_3$ (Refs.~\onlinecite{Marronnier_JPCLett2017,Marronnier_blackphases_ACSNano2018} for $\alpha$,$\beta$ and $\gamma$ and Ref.~\onlinecite{Roma_MSMSE2019} for $\delta$) and further relaxed with the mentioned cutoffs and {\bf k}-points.
The MAPbI$_3$ tetragonal structures were taken from Ref.~\onlinecite{Tong_2022}.
With our xc-functional some differences in the ordering of the six tetragonal polymorphs with respect to Ref.~\onlinecite{Tong_2022} arise.\cite{KMadaanThesis_2023} However, the most stable polymorph, which we refer to as tetragonal A, is the same in both cases.

The concept of ``polymorphism'' has been recently discussed for halide perovskites, in particular for their high temperature cubic phase,\cite{Zhao_2020}.  The meaning of polymorphism, here, is slightly different than the usual one: it indicates, for these materials, a phase which has long range disordered distortions whose average is the high simmetry (so-called ``monomorphous'') structure with a unit cell containing only a few atoms (5 for the inorganic perovskites, 12 for MAPbI$_3$).
The disordered structure in this case has lower internal energy at 0K than the high symmetry structure. 
In our work, we generated the polymorphous structure of MAPbI$_3$ using the same procedure proposed by Zunger and coworkers \cite{Zhao_2020} and a large supercell containing 32~fu (384 atoms, 2$\sqrt{2}\times$2$\sqrt{2}\times$4 of the 12 atoms unit cell) as explained in detail in Ref.~\onlinecite{KMadaanThesis_2023}. For the inorganic CsPbI$_3$ and CsPbBr$_3$ we used a different approach to generate a polymorphous structure: the special displacement method (SDM)\cite{Zacharias_SDM_PRR2020,Zacharias_PRB2023}, that allows us to obtain a dynamically stable structure ---no imaginary frequencies in the full phonon dispersion--- with a smaller 2$\times$2$\times$2 supercell.
  The energy gain from monomorphous to polymorphous structures was 90 meV/fu for MAPbI$_3$ (95 meV with $\Gamma$ sampling only), 147 meV/fu for CsPbI$_3$, and 105 meV/fu for CsPbBr$_3$. With respect to previously reported energy gains, ours are $\sim$25-30\% larger both for MAPbI$_3$\cite{Zhao_2020} and for the inorganics\cite{Zacharias_PRB2023,Zacharias_npj2023}, which probably depends mainly on the choice of the exchange correlation functional. All polymorphous structures were relaxed keeping the original cubic symmetry of the monomorphous one. For the inorganic compounds, we kept the equilibrium volume of the monomorphous structure. For MAPbI$_3$, the volume of the polymorphous structure was also optimized, obtaining a slightly (1.7\%) smaller volume than the monomorphous one.

  \subsection{Defect calculations}
  Positron trapping generally occurs in neutral or negatively charged vacancy defects. For this reason, we focused our study of defects on cation and lead vacancies, whose stable states are expected to be negatively charged. Iodine (or bromine) vacancies, conversely, are expected to be positively charged and, as such, not able to trap positrons. According to previous defect studies the expected charge states for MA and Pb vacancies are -1 and -2 respectively\cite{Yin_APL2014,Meggiolaro_EES2018}. Charged defects calculations are performed in the standard way, with a compensating uniform background of charge. When we mention absolute formation energy of defects, we apply monopole charge corrections.
  In the case of vacancies in polymorphous cubic MAPbI$_3$ we also consider the correction due to the average potential shift in a supercell with a defect, $\Delta \overline{V}$. We calculate it as the energy shift minimizing the difference of the electronic density of states of the pristine and defected supercell. We found that this contribution is negligible and we don't include it in the presented formation energies. 
  When we show formation energies we specify the reference atomic/molecular ($\mu_A$, A=Pb,I,MA) and  electronic ($\mu_e$) chemical potential. For atomic chemical potentials, we consider iodine-rich conditions when:
  \begin{equation}
    \mu_I=\frac{E(I_2)}{2} \qquad \mu_{Pb}=E(PbI_2)-2\mu_I \qquad \mu_{MA}=E(MAI)-\mu_I+H_f(MAPbI_3)
    \end{equation}
  and for lead-rich conditions:
  \begin{equation}
    \mu_{Pb}=E(Pb) \qquad \mu_I=\frac{E(PbI_2)-\mu_{Pb}}{2} \qquad \mu_{MA}=E(MAI)-\mu_I+H_f(MAPbI_3) ,
  \end{equation}
  where E(I$_2$), E(PbI$_2$), E(Pb), E(MAI) are the calculated energies of, respectively, the I$_2$ molecule, the PbI$_2$ hexagonal crystal, metallic FCC lead, the tetragonal MAI molecular crystal. H$_f$(MAPbI$_3$) is the formation enthalpy of the MAPbI$_3$ phase under study with respect to MAI and PbI$_2$, or H$_f$(MAPbI$_3$)=E(MAPbI$_3$)-E(MAI)-E(PbI$_2$). In our case H$_f$(MAPbI$_3$) amounts to -0.164 eV for the tetragonal phase (A) and to -0.106 eV for the polymorphous cubic phase.
  
  Otherwise, we mostly discuss differences between formation energies of defects in the context of the polymorphous structures. For the latter we introduced vacancies in every possible site of our large supercell and relaxed the structure to calculate the dispersion in formation energies and positron lifetimes. 

  The binding energy of a positron to a defect, E$_b$(e$^+$), is calculated as the energy difference between the energy of the positron in the pristine, undefected, crystal lattice and the energy of the positron in the vacancy. It is thus by definition a positive quantity when the positron is bound to the defect. To calculate it we take into account that the zero of the positron potential is arbitrary and can be somewhat shifted by a quantity $\Delta \overline{V}$(e$^+$) when we add a defect to the supercell. Its expression is thus: E$_b$(e$^+$)=$\epsilon_{lattice}$- $\epsilon_V$ -  $\Delta\overline{V}$(e$^+$), where $\epsilon_{lattice}$ and $\epsilon_V$ are energy eigenvalues of the positron in the pristine and defected supercell, respectively. To estimate such kinds of potential shifts, for electrons or for positrons, a common approach is to evaluate the macroscopic average of the potential\cite{Baldereschi_1988}, i.e., the integral of the positron potential V(e$^+$) over a unit cell located far from the defect in a region where this quantity as a function of position is flat. In our case, this approach is not viable for two reasons: first, the unit cell of the tetragonal phase is already fairly large; second, the perturbation introduced by a defect tends to induce distortions beyond the size of the supercell, which makes it impossible to find a region where the macroscopic average is flat.
  We chose to inspect the value of the positron potential in two specific different environments: on the line corresponding to the C-N bond of the methylammonium molecular ion, and in a interstitial region at equal distance between two iodine atoms. In both cases, without defects, the spread of minimum values does not exceed 0.004 eV for 8 and 16 fu supercells and 0.013 eV for the largest 32~fu supercells. When the supercell contains a defect, the values have a larger spread. To estimate the potential shift we consider the average of the positron potential at the minima along the C-N bond whose length is 2.82 bohr, or at the point in the middle of the line connecting two iodine atoms. We select iodine atoms whose distance is between 12.2 and 12.3 bohr. Interstitial regions and minima along C-N bonds give very similar results, although not identical. We finally choose the C-N bond which suffers a smaller spread. With this tool, we find potential energy shifts ranging from 0.2 eV for V$_{Pb}$ in the 8 fu tetragonal unit cell, to 0.03 eV for V$_{MA}$ in the large 32 fu supercell. For a given defect and supercell size the potential energy shift is almost identical for all choices of enhancement factors. Further details on the potential energy shifts are given in the Supplemental Material.\cite{SuppInfo}

  Voronoi volumes of the various pristine lattices and supercells containing vacancies were obtained by using the Freud python module,\cite{freud2020} version 0.1.2, with the option for periodic boundary conditions.

\section{Results}
\label{Results}

\subsection{Positron lifetimes in lattice: composition and polymorphism effect }
\label{StablePhases}

We present here a comparison of positron lifetimes calculated for the various phases of MAPbI$_3$, from the low temperature orthorhombic phase, to the five tetragonal polymorphs\cite{Tong_2022} and the high temperature cubic phase. Concerning the cubic phases, we consider both the high symmetry idealized small unit cell with only one formula unit, so called monomorphous, and the polymorphous supercell.\cite{Zhao_2020}
We add for comparison positron lifetimes that we calculated for two inorganic halide perovskites, CsPbI$_3$ and CsPbBr$_3$ in their various phases, including the polymorphous version of the high temperature cubic $\alpha$ phase. For the latter, the supercells are not as large as the one used for MAPbI$_3$. However, they should reliably represent the high temperature phase, given the method with which the structures were obtained (see section \ref{Methods}).
As shown in Table \ref{Tab:PALpristine}, we calculated positron lifetimes with the parameter free GGA enhancement factor proposed by Barbiellini and Kuriplach (B15-GGA)\cite{Barbiellini_2015}, both using DFT electronic densities and the crude approximation of a superposition of atomic densities. The latter is considered to allow the comparison with results obtained with the same approximation in the literature.\cite{Keeble_2021}

\begin{table}[ht]
\caption{Positron annihilation lifetimes in the lattice for the various phases of MAPbI$_3$ calculated both using a superposition of atomic densities or the self-consistent Kohn-Sham electron density. Lifetimes calculated in cubic CsPbI$_3$ and CsPbBr$_3$ are also shown for comparison. The enhancement factor used here is the one proposed in Ref. \onlinecite{Barbiellini_2015} (B15-GGA). }
\begin{tabular}{lcc} 
 \hline\hline
   Positron Host   & \multicolumn{2}{|c}{Positron Annihilation Lifetime (ps)}  \\ \hline
	& atomic densities & DFT densities \\    
 \hline
 MAPbI$_3$, Orthorhombic (O) &   289     &       297  \\
 MAPbI$_3$, Tetragonal (T$_A$) & 301          & 313  \\
 MAPbI$_3$, Tetragonal (T$_B$) & 301          & 313  \\
 MAPbI$_3$, Tetragonal (T$_C$) & 300          & 312  \\
 MAPbI$_3$, Tetragonal (T$_D$) & 301          & 314  \\
 MAPbI$_3$, Tetragonal (T$_E$) & 301          & 312  \\
MAPbI$_3$, Cubic, polymorphous (P) & 302      & 314  \\ %
MAPbI$_3$, Cubic, monomorphous (C) & 307     & 318 \\   %
$\delta$ CsPbI$_3$, orthorhombic & 285  & 292 \\
$\gamma$ CsPbI$_3$, orthorhombic & 308  & 313 \\
$\beta$ CsPbI$_3$, tetragonal & 309 & 313 \\
$\alpha$ CsPbI$_3$, cubic, polymorphous & 320 & 326 \\
$\alpha$ CsPbI$_3$, cubic, monomorphous & 319  & 322 \\
$\delta$ CsPbBr$_3$, orthorhombic & 269  & 277 \\
$\gamma$ CsPbBr$_3$, orthorhombic & 291   & 297  \\
$\beta$ CsPbBr$_3$, tetragonal & 289  & 293 \\
 CsPbBr$_3$, cubic, polymorphous &  300 & 306 \\
 CsPbBr$_3$, cubic, monomorphous & 299  & 303 \\
 \hline\hline
\end{tabular}
\label{Tab:PALpristine}
\end{table}

\begin{figure}
  \includegraphics[width=\columnwidth]{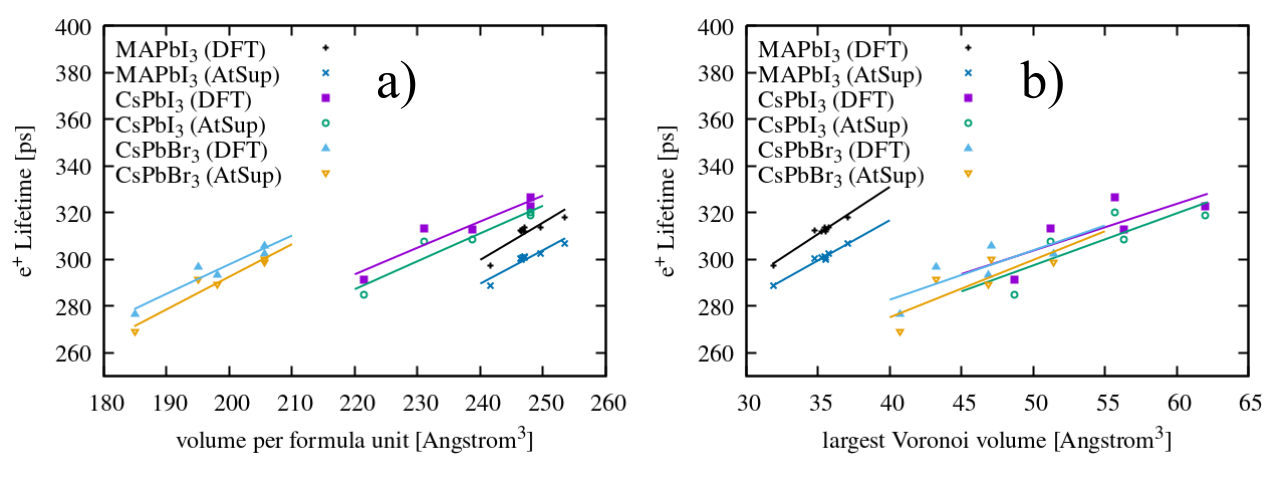}
	\caption{Calculated positron lifetimes for the pristine phases of MAPbI$_3$, CsPbBr$_3$ and CsPbI$_3$ $vs$ the volume per formula unit (a) or the largest Voronoi volume in the cell (b). Lines are linear regressions to the data. For each structure we show both positron lifetimes calculated with DFT electronic charge densities and with a superposition of atomic densities (AtSup).}
  \label{FigPALvsV} 
\end{figure}

The positron lifetime in the MAPbI$_3$ lattice is monotonically increasing with the volume per formula unit of the associated phase. This is indeed expected. For the inorganic counterparts CsPbI$_3$ and CsPbBr$_3$, the lifetime is comparatively higher for the same volume per formula unit, as can be seen from the data shown in figure \ref{FigPALvsV}a.
This difference is due to the different cation.
We chose to plot the data also as a function of the largest Voronoi volume of the crystal lattice (Fig. \ref{FigPALvsV}b).
Such a plot will allow a more thorough analysis when we will include the lifetimes in vacancies, in section \ref{Vacancies}. We note that the fit to the data of the two inorganic compounds, when performed  {\sl vs} the Voronoi volumes, gives essentially the same curve, clearly distinct with respect to the hybrid organic-inorganic perovskite.

The comparison in Table~\ref{Tab:PALpristine} between the positron lifetimes in the lattice calculated with a superposition of atomic densities and those including a fully self-consistent DFT electronic density deserves some further comments. The former are shifted to lower lifetimes by approximately 2 to 10 ps with respect to the latter. The shift calculated with B15-GGA is smaller for the inorganic compounds, $\sim$5~ps, and is larger for hybrid ones, from 10~ps in Table~\ref{Tab:PALpristine}  for the B15-GGA enhancement factor and up to $\sim$30~ps,  for the earlier GGA approach ---B95-GGA  one, (not shown in Table~\ref{Tab:PALpristine}).

The small decrease in positron lifetimes at 0K from the monomorphous to the polymorphous cubic lattices, 4.5~ps (Table \ref{Tab:PALpristine}), stems at least partly from the decrease in equilibrium volume and the difference in the pair distribution functions.
This hints towards the influence of temperature on the lifetime. Indeed, the pair distribution function of the polymorphous structure that we calculated here at 0K is in much better agreement with the experimental one\cite{Billinge_2016,Zhao_2020} at 350K for cubic MAPbI$_3$ than the monomorphous cubic.

At variance with MAPbI$_3$, we did not relax the equilibrium volume for polymorphous CsPbI$_3$ and CsPbBr$_3$. As a consequence, the calculated positron lifetime at 0K is slightly larger, 3.5~ps, for the polymorphous than for the high symmetry monomorphous lattice (DFT column in Table \ref{Tab:PALpristine}).  This increase in positron lifetime is similar to previous estimations of temperature effect on lifetimes in silicon estimated by only displacing atoms along phonon eigenvectors.\cite{Simula_2022}  This comparison highlights the fact that a careful estimation of temperature effects should include both atomic displacements, volume variations due to polymorphism but also to thermal expansion, i.e., one should minimize the full free energy of the system versus volume. 
 
\subsection{Positron lifetimes in positrons traps in MAPbI$_3$}
\label{Vacancies}
Positrons are delocalized in materials unless they encounter trapping centers, such as ionized acceptors or neutral vacancies, vacancy clusters, voids.

Iodine interstitials are expected to be ion-type ionized acceptors in MAPbI$_3$.  We calculated the lifetime for a 16 formula units supercell containing 193 atoms and a negatively charged iodine interstitial with the two B15-GGA and B95-GGA  considered in section~\ref{StablePhases} for the enhancement factor. The  lifetimes is shorter by 5~ps than the corresponding lattice one, and the binding energy is negligible ($<$0.03~eV). 

Here we focus then on vacancies in MAPbI$_3$. Lead and methylammonium vacancies, are mostly in charge states $-2$ and $-1$, respectively\cite{Yin_APL2014,Meggiolaro_EES2018}.  Iodine vacancies are expected to be stable only if positively charged, and thus unable to trap positrons.

For the comparison of the calculated lifetimes of a positron trapped in lead and methylammonium vacancies in MAPbI$_3$, we first focus on the most stable tetragonal polymorph, expected to be the stable phase at room temperature.

\subsubsection{Positron lifetimes in positrons traps in MAPbI$_3$: vacancies in the tetragonal phase}

\begin{table}
  \caption{Convergence with the supercell size of the calculated lifetimes ($\tau$) and binding energies (E$_b$) of a positron in lead and methylammonium vacancies, for the three GGA flavors of the enhancement factors as well as the Boronski-Nieminen LDA, taken as a reference. Cell sizes for tetragonal MAPbI$_3$ are labelled by the number of formula units (fu) they contain.}
  \begin{tabular}{lcccccccc}\hline\hline
    cell size &
    \multicolumn{4}{c}{V$_{MA}^{-}$} &
    \multicolumn{4}{|c}{V$_{Pb}^{--}$} \\ \hline
  $\tau$ [ps] & BN-LDA & B95-GGA & K14-GGA & B15-GGA  & BN-LDA  & B95-GGA & K14-GGA &  B15-GGA \\    \hline
    8 fu     &  360.5 &    548.7 & 401.4  &    405.1  & 292.4  & 399.6   & 328.9 &    328.8         \\ 
    16 fu    &  363.0 &    568.3 & 409.9  &  415.2       & 291.0 & 399.1 & 327.7 &   327.7       \\ 
    32 fu    &  367.2 &    589.7 & 418.7  &  425.7       & 290.7 & 398.8 & 327.4  &  327.3       \\ \hline 
  E$_b$ [eV] &  BN-LDA & B95-GGA & K14-GGA   & B15-GGA  & BN-LDA   & B95-GGA & K14-GGA &  B15-GGA \\    \hline
    8 fu     &  0.86 & 0.52  & 0.54  &  0.50     & 0.51  & 0.53 & 0.49  &  0.49 \\ 
    16 fu    &  0.84 & 0.50  & 0.52  &  0.48     & 0.33  & 0.35 & 0.32   &  0.32 \\  
    32 fu    &  0.91 & 0.53  & 0.56  &  0.52    &  0.30  & 0.30 & 0.29  & 0.29  \\ \hline\hline
  \end{tabular}
  \label{TabCellSizeConvergence}
\end{table}

We calculated the positron lifetime at 0K for tetragonal supercells of various sizes containing a vacancy. As seen in Table~\ref{TabCellSizeConvergence} the lifetime of the lead vacancy seems to be relatively well converged already with a supercell containing 16 fu, while the lifetime of the methylammonium vacancy, as well as binding energies of the positron to the vacancy are not yet fully converged even for our largest supercell, containing almost 400 atoms.
The binding energy is clearly larger for the methylammonium vacancy, than for the lead one with both the semilocal EPCF presented in Table~\ref{TabCellSizeConvergence}.
It is clear that, for MAPbI$_3$, the spread in lifetimes values betweeen the various approximations for the enhancement factor is very large, with huge variations in the case of the methylammonium vacancy. Even the difference between the two newest GGA approximations (which give very similar results) and the oldest B95-GGA is huge, $\sim$ 70~ps for the lead vacancy and more than 150~ps for the methylammonium vacancy.
These values are well beyond those found in other materials.\cite{Barbiellini_2015,Simula_2022}
We stress that the electron density used for each defect is exactly the same for all approximations.

 Concerning the methylammonium vacancy, the calculated lifetime is quite long, especially with the oldest GGA approximation~\cite{Barbiellini_1995} and the associated void, as measured from the largest Voronoi volume, relatively large. In such voids we expect regions of very low electronic density, where the screening of the interaction is strongly reduced, and significant density variations. In such cases non-local correlations neglected by semilocal approximations, may possibly be better described by a non-local approach like the WDA.
This approximation depends on the screening parameter Q whose the choice is somewhat problematic.

\begin{figure}
  \includegraphics[width=\columnwidth]{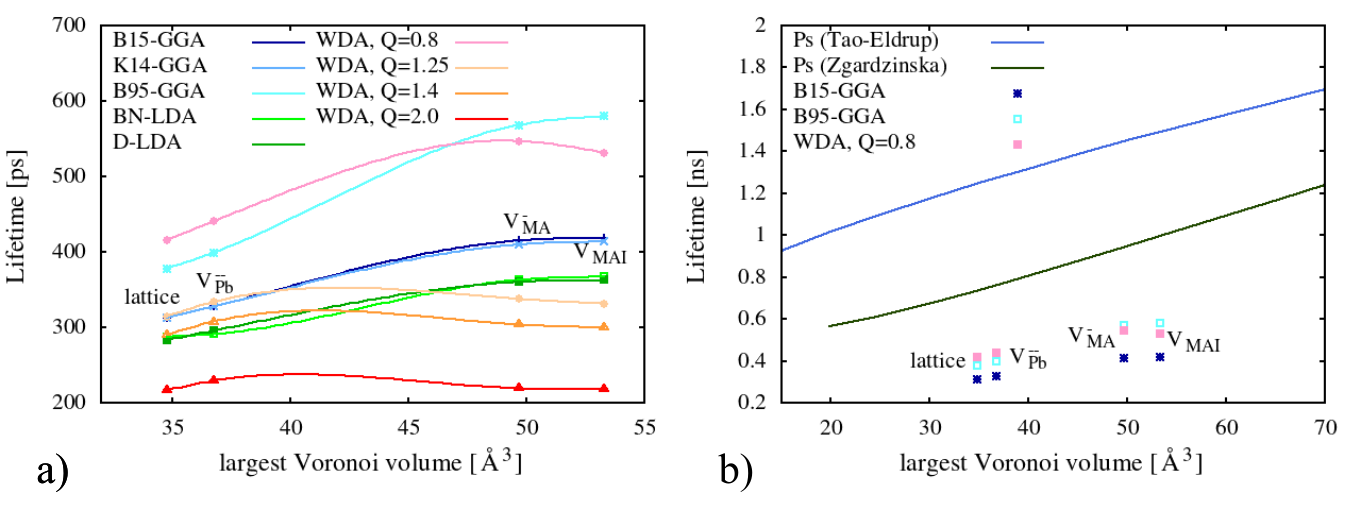}
  \caption{Comparison of lifetimes in tetragonal MAPbI$_3$ with and without vacancies for various enhancement factors. The charge states are -2 for the lead vacancy, -1 for the methylammonium one, and neutral for the MAI divacancy. Defects calculations were in 16 fu supercells. (a) Lifetimes are presented as a function of the volume of the void in which the vacancy is trapped (i.e., the largest Voronoi volume of the cell) for various approximations. The right panel (b), shows the comparison of three approximations (with the longest lifetimes) with ortho-positronium lifetime according to the Tao-Eldrup\cite{TaoEldrup_1972} model and the corrected version by Zgardzinska\cite{Zgardzinska_2015}, for the same void volumes.}
  
  \label{FigCompEnh}
  \end{figure}

Although the calculations of the lifetime for the large supercell containing 32 formula units is easily attainable with GGA or LDA functionals, it is prohibitively long with the WDA approximation. For this reason our comparison of different approximations for the lifetime of positrons trapped in vacancies makes use of the 16 fu supercell (Figure \ref{FigCompEnh}a). Together with the Pb and MA vacancies we add the neutral MA-I divacancy (V$_{MAI}$), which has a relatively strong binding energy\cite{KMadaanThesis_2023} and might be important in suppressing recombination of charge carriers.\cite{Qiao_2025}

The lifetime for the methylammonium vacancy is larger than 400~ps with the three GGA approximations (Table~\ref{TabCellSizeConvergence}). 
With this kind of long lifetimes, and large void spaces, then two questions arise: First, are the semilocal approximations able to capture the strong density variations associated with cases like this? Second, is it reasonable to compare the calculated lifetimes for free positrons to experimental situations in which, due to the large voids, the formation of positronium atom could be expected? Answering to the first question is, of course, a hard task, but our choice of testing a fully semilocal approximation (the WDA), which has been shown to provide useful insights in the case of surfaces,\cite{Shi_PositronCdSeWDA_PRL2018} is supposed to give a partial answer. To address the second question we show, in figure \ref{FigCompEnh}b, the expected lifetime of ortho-positronium as a function of the available void space according to the classical Tao-Eldrup model\cite{TaoEldrup_1972} and a recent corrected version of it.\cite{Zgardzinska_2015} For the comparison to our results we consider the spherical volume of the two mentioned models as equivalent to the Voronoi volume associated to a defect. It is clear from figure \ref{FigCompEnh}b that, according to those models, ortho-positronium annihilation lifetimes are much larger than those we have calculated. Thus the two annihilation modes can clearly be distinguished experimentally.

\begin{figure}
  \includegraphics[width=0.9\columnwidth]{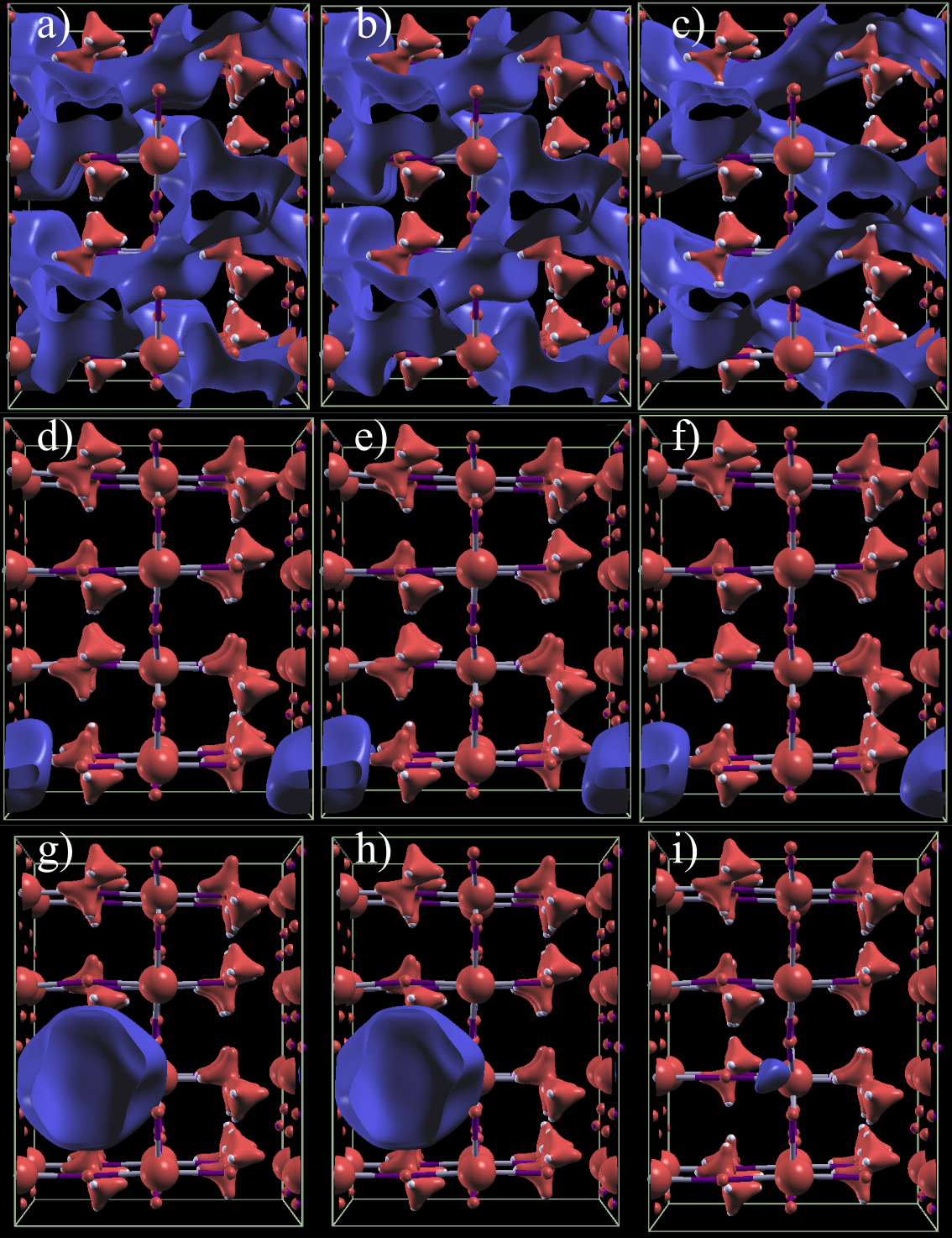}
  \caption{Comparison of electron (red) and positron (blue) charge density isosurfaces for the pristine tetragonal phase (a, b, c), a lead vacancy (d, e, f) and a methylammonium vacancy (g, h, i) for three different choices of the electron-positron correlation functional: GGA-B15\cite{Barbiellini_2015} (a, d, g), GGA-B95\cite{Barbiellini_1995} (b, e, h), and WDA\cite{Callewaert_2017} with Q=1.25 (c, f, i). Isovalues for the densities are all at 0.16 electrons/bohr$^3$, but positron densities have been scaled to be comparable to electron densities (scaling factors of 50, 10, and 4 for pristine phases, methylammonium vacancies, and lead vacancies, respectively).}
  \label{FigCompareVMAdensities}
\end{figure}

To better understand the positron behavior in the large voids associated with the methylammonium vacancy, we have applied the WDA approximation with different values of the screening charge Q for the positron lifetime calculations as shown in Figure~\ref{FigCompEnh}a.  An interesting point that can be inferred from Figure 2a is that by varying the values of the screening charge, Q, the positron annihilation lifetime in tetragonal MAPbI$_3$ lattice and lead vacancy calculated with the other approximations can be roughly reproduced by changing the Q values. For example, in Figure~\ref{FigCompEnh}a, the value Q=1.25 gives WDA values (orange curve) for those two quantum states that are similar to those calculated with the B15-GGA approximation (blue curve). For increasing Q values in Figure~\ref{FigCompEnh}a, Q=0.8, 1.25, 1.4, 2.0, the positron lifetimes clearly decrease for the three positron quantum states, lattice and  lead or methylammonium vacancy. However, the positron state associated to the methylammonium vacancy, exhibits stronger lifetime variations than the two other states. In addition, the comparison with the other approximations shows that the methylammonium vacancy has a different behavior from the other states. Those states,  depending on the Q values and approximation, are associated to lifetimes for which the variations are consistent whereas such a consistency is absent for the methylammonium vacancy. For example, for Q=0.8 in Figure~\ref{FigCompEnh}a, the lifetimes for the lattice and the lead vacancy are longer in the WDA approximation than within the B95-GGA, while they are shorter for the methylammonium vacancy. The specific behavior of the methylammoniun vacancy justifies the treatment within the WDA approximation. The limit of large screening charge, Q=2, would apply for Ps$^\textrm{-}$ ion formation in a {\sl homogeneous} electron gas at low densities close to those found in a vacancy site. However, in a perovskite material, ionic bonds and local dipoles would influence the stability of a Ps$^\textrm{-}$ so that it is impossible to predict its stability without a full many-body calculation taking into account the electron density of the solid. Our calculated free positron annihlation lifetime in vacancies is much lower than that expected for the Ps$^\textrm{-}$ (479~ps \cite{Frolov_1999,Fleischer_2006,Ceeh_2011}).

We inspected the calculated positron densities for supercells with and without vacancies obtained with the WDA (Q=1.25) and with two GGA approximations (see figure~\ref{FigCompareVMAdensities}). To present on the same figure electron and positron densities, the latter have been scaled by a factor mentioned in the legend of figure~\ref{FigCompareVMAdensities}. An alternative view, with the same scaling, is in the Supplemental Material.\cite{SuppInfo}
As can be seen in figure~\ref{FigCompareVMAdensities}, the shape of the isosurface for V$_{MA}$ differs significantly between the WDA and GGA results. It is thus clear in the case of the large void associated with the methylammonium vacancy that, whatever the most suitable value of the screening charge Q, the WDA approximation captures some physical features that are overlooked by all the other semilocal functionals.  This is corroborated by inspection of the profile of the total positron potential for vacancies (see Supplemental Material section D and figure S5). The  GGAs and LDA approximations present a minimum at the vacancy sites, while the WDA shows a slight maximum. Such behaviour could probably not be reproduced by an improved fitting of the $\alpha$ parameter of the GGA.

\begin{figure}
  \includegraphics[width=0.7\columnwidth]{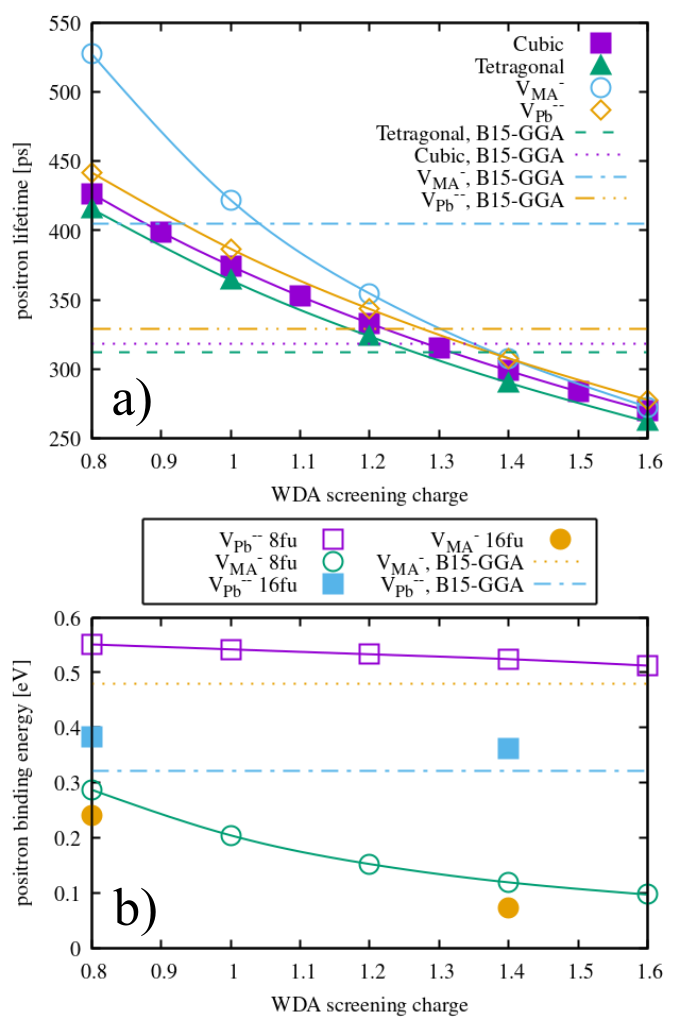}
  \caption{Influence of the chosen value of the WDA screening charge Q on the calculated positron lifetimes (panel a) and binding energies (panel b). Vacancies are all in the tetragonal phase. The corresponding values obtained with the B15-GGA enhancement factor are shown with dashed/dotted lines. In panel a) results were obtained with 8fu supercells.}
  \label{FigQeffect}
  \end{figure}

To further investigate the influence of the choice of the screening charge Q, we performed calculations at various Q values and we show the results for the lifetime and the associated binding energy in figure \ref{FigQeffect}. Of course, for each calculation of the binding energy, both eigenvalues and positron potential alignment correction have been obtained with the same value of Q.
The curves in panel a) clearly confirm that the behavior of V$_{MA}$ is peculiar, with a stronger dependence to the screening charge with respect to the pristine tetragonal cell or the lead vacancy. Panel b) shows a similar trend, with a stronger variation of the binding energy with the screening parameter Q for V$_{MA}$ than for V$_{Pb}$. However, the most striking feature is probably that, with the WDA approximation, the positron is much more strongly bound to V$_{PB}$ than to V$_{MA}$: this is another clear qualitative difference between the WDA and the other approximations, whatever the choice of the value of the screening charge Q.
This inversion in binding energy between the B15-GGA and the WDAs can be attributed to the peculiar behavior of the positron density and the total positron potential, which are qualitatively different between B15-GGA and WDA for V$_{MA}$, while they are closer in shape for V$_{Pb}$ (see Fig.~~\ref{FigCompareVMAdensities} for the density and Fig.~S4 for the potential).

\subsubsection{Positron lifetimes in positrons traps in MAPbI$_3$: vacancies in the polymorphous cubic phase}
\label{PolymorphousCubic}

Polymorphism in the cubic phase of MAPbI$_3$ manifests itself in a variety of local environments for point defects and, thus, also for the positrons trapped in them. To the best of our knowledge, the distribution of formation energies of defects induced by polymorphism still need to be investigated in detail. While approaches more advanced than ours, and much more computationally demanding, to defect energies have been employed (including, e.g., both hybrid functionals and spin-orbit coupling), we use here, as described in section \ref{Methods}, a semilocal van der Waals functional, without spin orbit coupling, which we expect to provide a useful estimation of the spread of formation energies due to polymorphic distortions.

After calculating the formation energy of vacancies inserted in any possible site in our polymorphous model with 32 formula units, we analysed the formation energies
and the positron lifetimes in Table~\ref{Tab:Poly_spreads}. The table contains also the average potential shift $\Delta \overline{V}$ (see section~\ref{Methods} for details).
The spread of $\Delta \overline{V}$ is much smaller than the standard deviation on formation energies.

\begin{table}
  \caption{Average and standard deviation (in parentheses) of various quantities in polymorphous cubic MAPbI$_3$ for three different vacancy defects: defect formation energies (E$_f$), positron potential shift ($\Delta \overline{V}$), positron binding energy (E$_b$), and positron lifetime ($\tau$).   Lifetimes and binding energies are calculated with the B15-GGA enhanceent factor. Shifts of
    the average potential ($\Delta \overline{V}$) are calculated with respect to the pristine polymorphous supercell. The Fermi level, for the calculation of formation energies, is set at the top of the valnce band.}
  \label{Tab:Poly_spreads}
 \begin{tabular}{lcccc}\hline\hline
	  Defect    &    E$_f^{I-rich}$ [eV]  & $\Delta \overline{V}$ [eV] & E$_b$(e$^+$) [eV] & $\tau$ [ps] \\ \hline
    V$_{MA}^-$   &      1.02 (0.016)    & -0.063 (7$\times$10$^{-4}$)   & 0.45 (0.012)       &   417.4 (1.96) \\    
   V$_{Pb}^{--}$ &  0.64 (0.018)       & -0.079 (1.2$\times$10$^{-3}$) & 0.20 (0.007)        & 322.7 (0.82) \\
    V$_I^+$     &  0.51 (0.064)       &  -0.029 (3.2$\times$10$^{-3}$) & ---               &    ---- \\ \hline
  \end{tabular}
\end{table}

For all the quantities shown in Table~\ref{Tab:Poly_spreads}, the standard deviation is fairly small.

\begin{figure}
  \includegraphics[width=0.95\columnwidth]{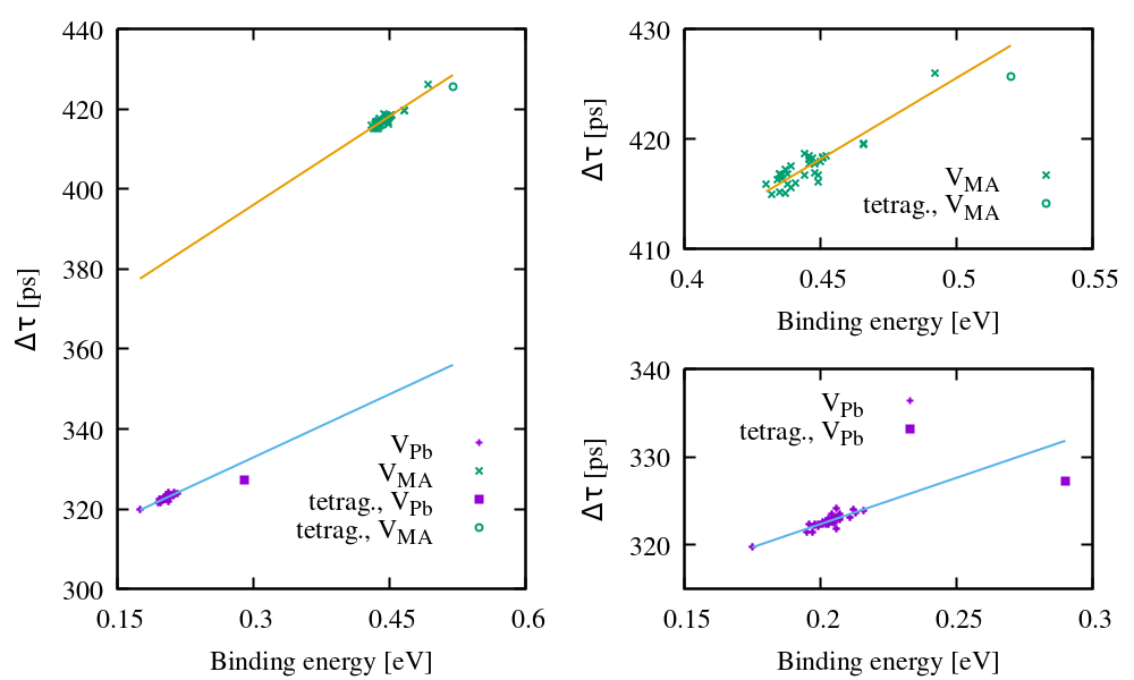}
  \caption{Positron lifetime in lead and methylammonium vacancies, with the B15-GGA approximation, as a function of the binding energy of the positron to the vacancy, in polymorphous cubic and tetragonal MAPbI$_3$. The lines are linear fits to the vacancies in the polymorphous cubic only. The two panels on the right are zooms on the relevant regions of the left panel.}
  \label{Fig:Tau_vs_Eb}
\end{figure}

We checked whether the small spread of the positron lifetime shown in Table~\ref{Tab:Poly_spreads} stems from  the Voronoi volume associated with each defect site available in the polymorphous structure. In this case, where the variations of the Voronoi volumes remain small, $<$3.8\%, there is no correlation with the small lifetime variation, $<$2.6\%. Conversely, we found a correlation between the lifetime and the binding energy of the positron to the vacancy.
In figure \ref{Fig:Tau_vs_Eb} we show the lifetime {\sl vs} the binding energy calculated for  vacancies in the polymorphous cubic cell. The lines are linear regressions to these data. For comparison the lifetime for vacancies in the tetragonal phase are also shown, but not included in the fit.
The positron potential shift correction was calculated for only one vacancy of each type in the polymorphous supercell, and applied to all the vacancies of the same type, V$_{MA}^-$ or V$_{Pb}^{--}$.
The plot shows a roughly linear relationship  between binding energy and lifetime.

\subsection{Discussion}
\label{Discussion}

Our calculations show that various approximations for the electron-positron correlation can result in positron lifetimes that, for the methylammonium vacancy, vary by more than 200~ps (Figure \ref{FigCompEnh}), which is huge. In this discussion, we first compare our calculations to those reported previously. We then compare them to experimental values to determine whether one approximation with respect to others can be singled out as yielding values that are more consistent with the experimental ones. Such a consistency is a criterion for selecting the best approximation with which to describe the electron-positron correlation in halide organic-inorganic perovskites.

Concerning CsPbBr$_3$ our lifetime results for the crystal lattice of the different phases are clearly lower than the theoretical prediction by Musiienko {\sl et al.}, 355~ps. This lifetime is calculated according to the standard approach, the same as ours, and the BN-LDA approximation. The differences with our calculation might stem from the phase (presumably monomorphous cubic) and other technical parameters (basis set, pseudopotentials), for the DFT calculation of the electronic density, which are not specified. In our case, using the BN enhancement factor for the monomorphous cubic phase gives a much lower lifetime of 270~ps.

The theoretical results reported by Keeble {\sl et al.}\cite{Keeble_2021} for MAPI employ three different computational schemes: 1) with atomic superposition of atomic densities; 2) with a self-consistent DFT scheme including the density of electrons and the positron; 3) with a self-consistent DFT scheme as 2) plus the relaxation of the atomic positions induced by the presence of the positron. We can compare the results of 1) with our calculations with superposition of atomic densities and those of 2) with our calculations with DFT densities.

For the first comparison, we find a very similar value for calculations of the positron lifetime in the tetragonal MAPbI$_3$ lattice using electronic density from atomic superpositions and the B95-GGA enhancement factor, 342~ps in our case {\sl vs} 353~ps in Ref.\onlinecite{Keeble_2021}.
However, in our case, for relaxed supercell sizes with 32~fu, the calculated lifetimes for the V$_{MA}^-$ and V$_{Pb}^{--}$ vacancies are 345 and 365 ps, definitely smaller than the values of Keeble {\sl et al.}. It is difficult to pinpoint the reason for the better agreement for the lattice than for vacancies. For the lattice, the equilibrium volume is 0.8\% smaller than their's. For vacancies, the difference can likely  be ascribed to the differences in the choice of the supercells and the details of the relaxation of the atomic positions.

  Concerning the second comparison, we recall that
  in Ref.~\onlinecite{Keeble_2021} the calculation of the positron lifetimes takes into account self-consistently the modification of the electron density due to the presence of the positron, and vice-versa and, apparently, using Boronski-Nieminen (BN) e$^+$-e$^-$ correlation coupled with the gradient corrections
of B95-GGA  for determining the enhancement factor.\cite{Wiktor_2015,Wiktor-privcomm} This is not the standard scheme that we use, where the modification of the electron density due to the positron is taken into account through the enhancement factor. For this reason it is not straightforward to compare our DFT based results to the DFT ones in  Ref.~\onlinecite{Keeble_2021}. While the full self-consistency between electron and positron densities ---and even positron induced relaxation of atomic positions--- may play a role, it is unclear what are the contributions coming from electron-positron self-consistency and what is the consequence of using Boronski-Nieminen (BN) e$^+$-e$^-$ correlation coupled with the gradient corrections of B95-GGA, which is, apparently, the choice that was made to obtain the mentioned results.
  It is important to note, that the fully self-consistent calculations in  Ref.~\onlinecite{Keeble_2021}, are performed taking the explicit zero-positron density limits of the correlation potentials and the enhancement factor,\cite{Gilgien_1994}
  a detail not mentioned in the publication itself.\cite{Wiktor-privcomm}
  This method has been shown to localize the positron stronger than the proper use of finite positron density.\cite{PuskaSeitsonenNieminen_1995}
  Our results obtained with the BN enhancement factor are 280, 291, 363~ps for the pristine tetragonal, lead and methylammonium vacancy respectively, thus much lower than 342, 360, 414~ps reported in the mentioned paper, while our lifetimes obtained with B95-GGA are larger (389, 398 and 568~ps, respectively).

  It is thus clear that further progress on the reliability of the various possible approximations for the calculation of the lifetimes is needed for this class of materials. It is possible that further combined tests on ionic materials, in comparison to covalent or partly covalent semiconductors, for pristine bulk and vacancies, could help clarify the issue, especially for materials for which it is possible to estimate the concentration of the main point defects present by different techniques. Revisiting the prediction of positron annihilation lifetimes and momentum densities in alkali halides might be worthwhile.

  Concerning the WDA approximation, which is a promising approach for theoretical studies of
  free positron annihilation in large voids and at surfaces, the dependence on the screening charge Q needs to be further investigated. To this goal, it would be interesting to revisit positron lifetimes and momentum densities in several materials with different bonding (metallic, covalent, ionic, weak bonds like hydrogen bonds and van der Waals), in order to ground the choice of Q on the peculiarities of the electron density and the electronic screening.\cite{Barbiellini_2025}
  We can nevertheless say that, in the case of MAPbI$_3$, Q values above 1.4, for which the positron lifetime in lead vacancies is equal or larger than the corresponding lifetime in methylammonium vacancies (Figure~\ref{FigQeffect}), appear as an unphysical choice, given the substantial difference in the associated Voronoi volumes.

  It is rather difficult to draw conclusions from the comparison between calculated and experimental reported values of the positron lifetimes, both for positron delocalised in the lattice and localised in vacancies. On the one hand, different theoretical approximations can give fairly different results and, on the other hand, we cannot unequivocally assign the dominant measured lifetime components either to the lattice or to any specific vacancy.
Let us list, nevertheless, the main experimentally determined lifetimes for hybrid or inorganic lead halide perovskites.
  Dhar {\sl et al.} report, as experimental positron decay lifetimes, values of 326\cite{Dhar_JPCC2017} and 333-335\cite{Dhar_JPCL2017}~ps for tetragonal MAPbI$_3$ (at room temperature) and 309~ps for the cubic phase (at 350~K). They attribute it to cation vacancies. The experimental values reported by Keeble,\cite{Keeble_2021} depending on the chosen sample, span a range from 367 to 375~ps, again for tetragonal MAPbI$_3$.
  The extrema of the range are in excellent agreement with the positron lifetimes identified as arising from two different quantum states by some of the present authors.\cite{CCa,CCb}
  If the lower value is to be attributed to the pristine lattice, then the lifetimes calculated with the B95-GGA, or WDA with Q$\simeq$1 (Table~\ref{Tab:PALpristine} and Figure~\ref{FigCompEnh}, light blue curve)
  are in better agreement. However, if both lifetimes represent mostly annihilation in vacancies, then the approximations giving lower lifetimes, e.g., B15-GGA or K14-GGA (Figure~\ref{FigCompEnh} darker blue curves) and WDA with larger screening charge, would be more appropriate. Actually, we cannot firmly exclude any of the two hypotheses.
  This comparison shows also that, in the second case, the calculated positron lifetime in tetragonal MAPbI$_3$ lattice is expected to be 313~ps. This value is much lower than the value 342~ps determined in Keeble {\sl et al.}\cite{Keeble_2021} by applying the trapping model with a single type of defects in order to analyse the decomposition of the measured lifetime spectra.
  This questions whether a specific Electron Positron Correlation Functional  scheme has to be developed to take into account the covalent character of the bonding in the organic cations.

\begin{figure}
    \includegraphics[width=0.95\columnwidth]{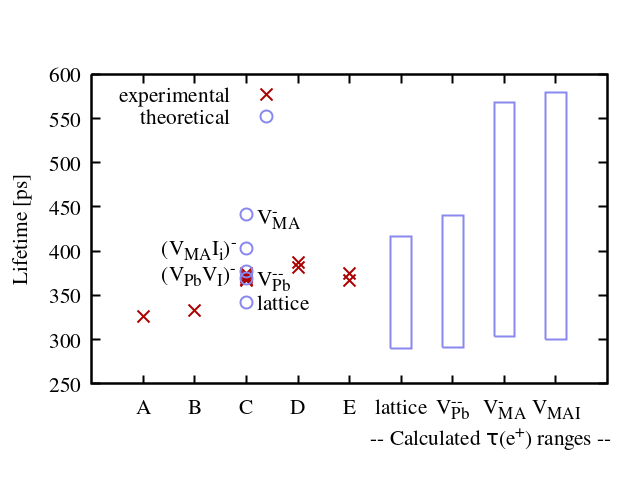}
    \caption{A summary of available experimental and theoretical positron lifetimes determined for MAPbI$_3$. A$\equiv$Ref. \onlinecite{Dhar_JPCC2017}, B$\equiv$Ref. \onlinecite{Dhar_JPCL2017}, C$\equiv$Ref. \onlinecite{Keeble_2021}, D$\equiv$Ref. \onlinecite{Cai_2025}, E$\equiv$Ref. \onlinecite{CCa}. The rectangles indicate the ranges of values obtained in this work by varying the electron positron correlation functional (using the data in table \ref{TabComparison}). The positron states of the calculated lifetimes from Ref.~\onlinecite{Keeble_2021} are indicated near the respective empty circles.}
    \label{FigComparison}
  \end{figure}

  To summarize the comparison of experimental and theoretical positron lifetime data for MAPbI$_3$, we collected the them in table~\ref{TabComparison} and, graphically, in figure~\ref{FigComparison}, which confirm that a firm assignment of positron lifetime to a given defect or to the lattice based on this comparison is presently unreliable.

  \begin{table}
    \caption{Theoretical positron lifetimes calculated in this work with various EPCF (16 fu supercell) in tetragonal MAPbI$_3$, compared to various experimentally determined lifetimes by various authors, using $^{22}$Na sources or slow positron beams. Units are ps.}
\begin{tabular}{lcccccccc} \hline\hline
\multicolumn{9}{c}{Theory (this work)}  \\ \hline
positron state & B15-GGA & K15-GGA & B95-GGA & BN-LDA & D-LDA & WDA Q=0.8 & WDA Q=1.25 & WDA Q=1.4 \\  \hline
pristine & 313 & 313 & 378 & 288 & 283 & 416 & 315 & 290 \\    %
V$_{Pb}^{--}$ & 328 & 328 & 399 & 291 & 296 & 441 & 334 & 308 \\   %
V$_{MA}^-$ & 415 & 410 & 568 & 363 & 360 & 547 & 338 & 304 \\  %
V$_{MAI}$ & 419 & 414 & 580 & 367 & 363 & 532 & 332 & 300 \\   %
\multicolumn{9}{c}{Experiment} \\ \hline
Reference & \multicolumn{3}{l}{ samples probed with $^{22}$Na $\beta^+$} & \multicolumn{5}{l}{samples probed with e$^+$ beam} \\ \hline
Ref. \onlinecite{Dhar_JPCC2017} & 326 & & & & & & & \\
Ref. \onlinecite{Dhar_JPCL2017} & 333 & & & & & & & \\
Ref. \onlinecite{Keeble_2021}  & & & & 367 & 368 & 370 & 371 & 375 \\
Ref. \onlinecite{Cai_2025}   & & & & 382 & 387 & & \\
Ref. \onlinecite{CCa} & & & & 367 & 375  & & & \\ \hline\hline
\end{tabular}

    \label{TabComparison}
    \end{table}

  The results by Musiienko {\sl et al.} are for bromides, CsPbBr$_3$ and MAPbBr$_3$, with lifetimes of 350 and 343~ps respectively.   For the $\gamma$-phase, which should be the stable one at room temperature, our calculated results with B15-GGA are clearly lower than the measured ones, which would suggest that maybe B95-GGA would be more suitable. According to Ni {\sl et al},\cite{Ni_2024} who find a dominant component of 344~ps for MAPbBr$_3$, such lifetime would mainly represent lead vacancies, given that their calculated value for the pristine lattice is lower (312~ps).

 As a general remark concerning the approximations giving the longest lifetimes (like B95-GGA), the question arises whether such values do correspond to situations where positronium could be formed, which cannot be described by our theoretical framework. The comparison with Voronoi volumes as shown in figure \ref{FigCompEnh} clearly tells us that, if positronium is formed, the lifetime is expected to be clearly longer than the largest calculated one. Then the answer can be only in the analysis of experimental results. On this point a short reminder about the experimental findings is perhaps worthwhile.

 Decay positron components with lifetimes longer than 450-500~ps and with low intensities have to be considered with care for experiments using $\beta^+$ radioactive ${}^{22}$Na salt as positron source. They may arise from the ${}^{22}$Na salt itself and/or from the material on which the radioactive source is deposited as mentioned in Ref.~\onlinecite{Krause_1999}. For example, 
 this is the case for the two papers by Dhar {\sl et al.}\cite{Dhar_JPCC2017,Dhar_JPCL2017} where, in MAPbI$_3$ pellets or single crystals,  the longer decay lifetime components have decay lifetimes between 1.050 and 1.275~ns with intensities, 8-9\%. 
The authors
attributed them to annihilation in large voids and surfaces present in their pellets or single crystals. However, there is no mention for the procedure used for source corrections in the analysis of the spectra.
Musiienko {\sl et al.}\cite{Musiienko_2022}, finds a long component, 1.5~ns lifetime with 1\% intensity, that is indeed attributed to positronium annihilation in the mylar foil sealing the $\beta^+$ ${}^{22}$Na radioactive salt.
For near surface in MAPbI$_3$, Keeble {\sl et al.}\cite{Keeble_2021} report the existence of long lifetime components with a large decay lifetime longer than 500~ps and intensities depending on the type of the films or single crystals. The components have intensities that are determined with statistical deviations in the range 13-166 \%. Such statistical deviations  suggest possible artefacts in the decompositions of the measured lifetime spectra that may depend on the  background noise.

  To conclude this discussion we comment on the correlation we have shown between calculated
  lifetimes and the largest Voronoi volumes associated to the positron quantum states for lattice and vacancies in MAPbI$_3$ and to the lattice in CsPbI$_3$ and CsPbBr$_3$.
 Such a correlation seems to be a better tool than that between lifetimes and theoretical densities to give rough estimations of lifetimes in the lattice. It seems also to work for vacancy defects in MAPbI$_3$.

\section{Conclusions}
\label{Conclu}

In this paper we have used a standard approach, based on first principles calculations, to the simulation of positron lifetime in methylammonium lead iodide with and without vacancies. We also compared, for annihilation in the lattice, positron lifetimes for various phases of  the inorganic lead halide perovskites CsPbI$_3$ and CsPbBr$_3$. We tested several approximations for the electron-positron correlation potential, including a fully non-local approach based on the weighted density approximation (WDA). Detailed results on vacancies in the tetragonal, room temperature, phase show that the discrepancies between the various approximations can be quite large and that the comparison with available experimental data does not allow to clearly single out a preferential approach.
We also analyzed binding energies of the positron to the vacancies; methylammonium vacancies show larger binding energy than lead ones for all semilocal approximations, but the opposite is true with the WDA.

We investigated several phases of MAPbI$_3$, including the polymorphic high temperature cubic phase; for the latter, which needs the use of very large supercells, we have studied the effect of local disorder on defect properties, estimating the standard deviation of vacancy formation energy, binding energies, lifetimes, and the position of the valence band edge.
Our calculated formation energies and positron lifetimes, show a relatively narrow distribution. Our approach could be applied to many other properties in polymorphous phases of halide perovskites.

We systematically compare our lifetime results with the available size of the voids available in the structure, estimated through Voronoi volumes; in many cases we find a linear relationship between them.

Further work, in tight connection with experimental studies, is needed to increase the capabilities of this technique to distinguish different kinds of defects in these peculiar materials, suggesting also revisiting positron lifetime annihilation predictions in simpler ionic materials.

\section*{Acknowledgments}
This study was funded by the French National Agency for Research (ANR) via the TRAPPER project n.19-CE05-0040.
This work was granted access to the HPC resources of TGCC and IDRIS under the allocations 2025-A0170906018, 2024-A0150906018 and 2023-A0130906018 made by GENCI.

\bibliography{PASpero}

\end{document}

\title{Supplemental Material to ''Challenges in predicting positron annihilation lifetimes in lead halide perovskites: correlation functionals and polymorphism.''}

\author{Kajal Madaan}
\affiliation{Université Paris-Saclay, CEA, Service de Recherches en Corrosion et Comportement des Matériaux, SRMP, 91191 Gif sur Yvette, France}
\author{Guido Roma}
\affiliation{Université Paris-Saclay, CEA, Service de Recherches en Corrosion et Comportement des Matériaux, SRMP, 91191 Gif sur Yvette, France}
\email{guido.roma@cea.fr}
\author{Jasurbek Gulomov}
\affiliation{Université Paris-Saclay, CEA, Service de Recherches en Corrosion et Comportement des Matériaux, SRMP, 91191 Gif sur Yvette, France}
\author{Pascal Pochet}
\affiliation{Department of Physics, IriG, Univ. Grenoble-Alpes and CEA, Grenoble,France}
\author{Catherine Corbel}
\affiliation{CEA/DRF/IRAMIS, University Paris-Saclay, Gif-sur-Yvette, France}
\author{Ilja Makkonen}
\affiliation{Department of Physics, University of Helsinki, P.O. Box 43, FI-00014 University of Helsinki, Helsinki, Finland}

\maketitle

\subsection{Brief review of experimental positron lifetimes}

\begin{figure}

  \include{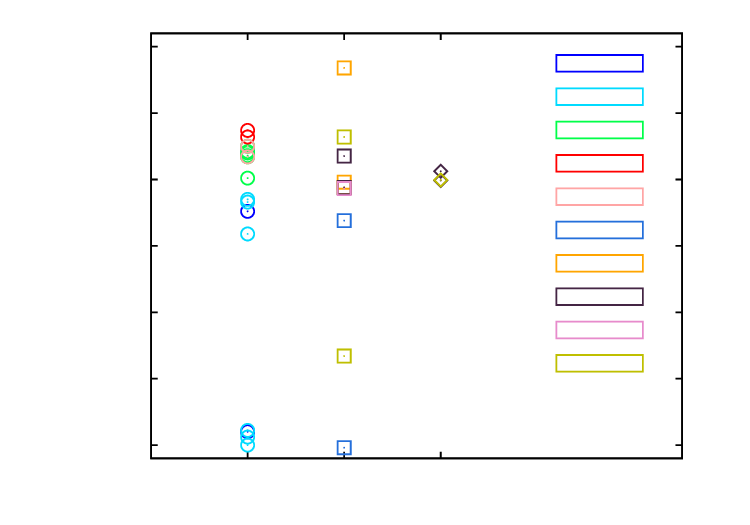}

  \caption{Measured positron lifetimes in various samples of three lead halide perovskites materials, obtained with different positron sources and experimental settings.}
\label{FigLiteratureReview}
  
  \end{figure}

In this section we will briefly comment on recent experimental determinations of positron lifetimes in halide perovskites; data have been published for MAPbI$_3$ \cite{Dhar_JPCC2017,Dhar_JPCL2017,Dhar_2018,Keeble_2021,Cai_2025}, MAPbBr$_3$\cite{Musiienko_2022,Moshat_2023,Ni_2024,Schmidt_2025} and CsPbBr$_3$\cite{Musiienko_2022}. We add some unpublished results for the three materials.\cite{CCa,CCb}
As reminded in the introduction of the main paper, 
lifetimes can be measured either deep within the bulk or at varying depth from the surface of the material, depending on whether a $\beta^+$ radioactive ${}^{22}$Na source or a slow positron beam is used. A global graphical summary of measured lifetimes is presented in Fig.~\ref{FigLiteratureReview}.

A few experimental data are available for tetragonal MAPbI$_3$ at 300~K. Dhar {\sl et al.} report positron decay lifetimes
measured deep inside tetragonal MAPbI$_3$\cite{Dhar_JPCC2017,Dhar_JPCL2017,Dhar_2018}. They measure short
values of 150$\pm$3 ps and long values of 333$\pm$3 ps at 300~K in single crystals\cite{Dhar_JPCC2017,Dhar_JPCL2017}. For
polycrystalline pellets obtained after mechano-synthesis \cite{Dhar_JPCC2017,Dhar_JPCL2017,Dhar_2018},
the values at 300~K depend on the grinding process and post-treatement: 156$\pm$3, 161$\pm$3 ps for the shorter positron
decay lifetimes and 309$\pm$5, 326$\pm$3, 335$\pm$3, 351$\pm$5 ps for the longer ones.
Near the surface of tetragonal MAPbI$_3$ for a mean depth of $\sim$0.81$\mu$m (16~keV positron energy beam), Keeble {\sl et al.}\cite{Keeble_2021} report shorter values varying from 42$\pm$24 to 119$\pm$45~ps and longer values varying from 367$\pm$1 to
375$\pm$1~ps for solution grown single crystals or spin-coated polycrystalline films at 300~K.
The experimental results obtained by %
Cai {\sl et al}.\cite{Cai_2025,CCa} show that the positron decay lifetimes at~300 K in polycrystalline pellets of tetragonal MAPbI$_3$ vary with depth,
(0-0.384 $\mu$m) below the surface. The unique decay lifetime resolved at 300~K in the positron lifetime
decay spectra go through a maximum as a function of depth and vary by 5~ps in the range 
of 382.0$\pm$.1 to 387.0$\pm$.2~ps depending on the grinding process.
In addition, by correlating the depth profile of the positron decay lifetime with the low and high momentum fractions of the e$^+$-e$^-$ annihilating pairs, the present authors have been able to identify two distinct positron quantum states associated to the positron annihilation lifetime, 367$\pm$1 and 375(1)~ps.\cite{CCa}

A few experimental data for other compositions than MAPbI$_3$ have been published. For the cubic phase of MAPbBr$_3$, deep within polycrystalline pellets obtained after mechano-synthesis, Moshat {\sl et al.}\cite{Moshat_2023} report short values of 148$\pm$5~ps and longer ones of 319$\pm$5~ps at 300K. For the cubic phase at 300 K and deep within the bulk of solution grown MAPbBr$_3$ single crystals, the positron decay lifetimes vary depending on the authors.
Musiienko {\sl et al.}\cite{Musiienko_2022} measure a lifetime of 343.2$\pm$.3~ps. Schmidt {\sl et al}\cite{Schmidt_2025} recently reported short values of 217$\pm$8~ps and longer values of 382$\pm$8~ps, independent of whether the measurements are performed in vacuum or in air.
For near-surface annihilation in the cubic phase of solution grown MAPbBr$_3$ single crystals at a mean depth of $\sim$500 nm (6 keV positron beam), Ni {\sl et al.} measure positron decay lifetimes depending on the stoichiometry of the precursor solutions.\cite{Ni_2024} For the MABr to PbBr mole ratios of 1:1, they report a short value of 344$\pm$5~ps and a longer one of 434$\pm$24~ps. 
For the MABr to PbBr mole ratio of 2:1, the dominant lifetime has a value 348$\pm$1~ps, slightly
longer by 4~ps than the short lifetime for the ratio 1:1. The single crystals obtained after addition of FABr to the 2:1 precursor solution give rise to a lifetime of 346$\pm$1~ps, only 2~ps lower than the value measured in the 2:1 precursor solution.

Some of the present authors have experimentally determined that the positron decay lifetimes at 300 K in MAPbBr$_3$ single crystals and spin coated polycrystalline thin films vary with increasing depth below the surface.\cite{CCb}  The decay lifetimes resolved at 300~K decrease by $\sim$23~ps, from 367.7$\pm$0.3~ps to 344.0$\pm$0.2~ps, in the single crystals as the mean positron implantation depth increases from $\sim$3  to 411~nm.  The lifetimes vary by~$\sim$3~ps and go through a minimum value of 345.2$\pm$0.2~ps as the mean positron implantation depth increases from $\sim$3 to 136~nm in the thin polycrystalline films.  In CsPbBr$_3$ thick ($>$10 $\mu$m) polycrystalline films, the decay lifetime resolved at 300~K vary by $\sim$7~ps and go through a minimum value of 349.2$\pm$0.2~ps as the mean depth increases from $\sim$2.7 to 329~nm.

\subsection{Localisation of the positron density with different enhancement factors}
\begin{figure}[h]
\includegraphics[width=0.9\columnwidth]{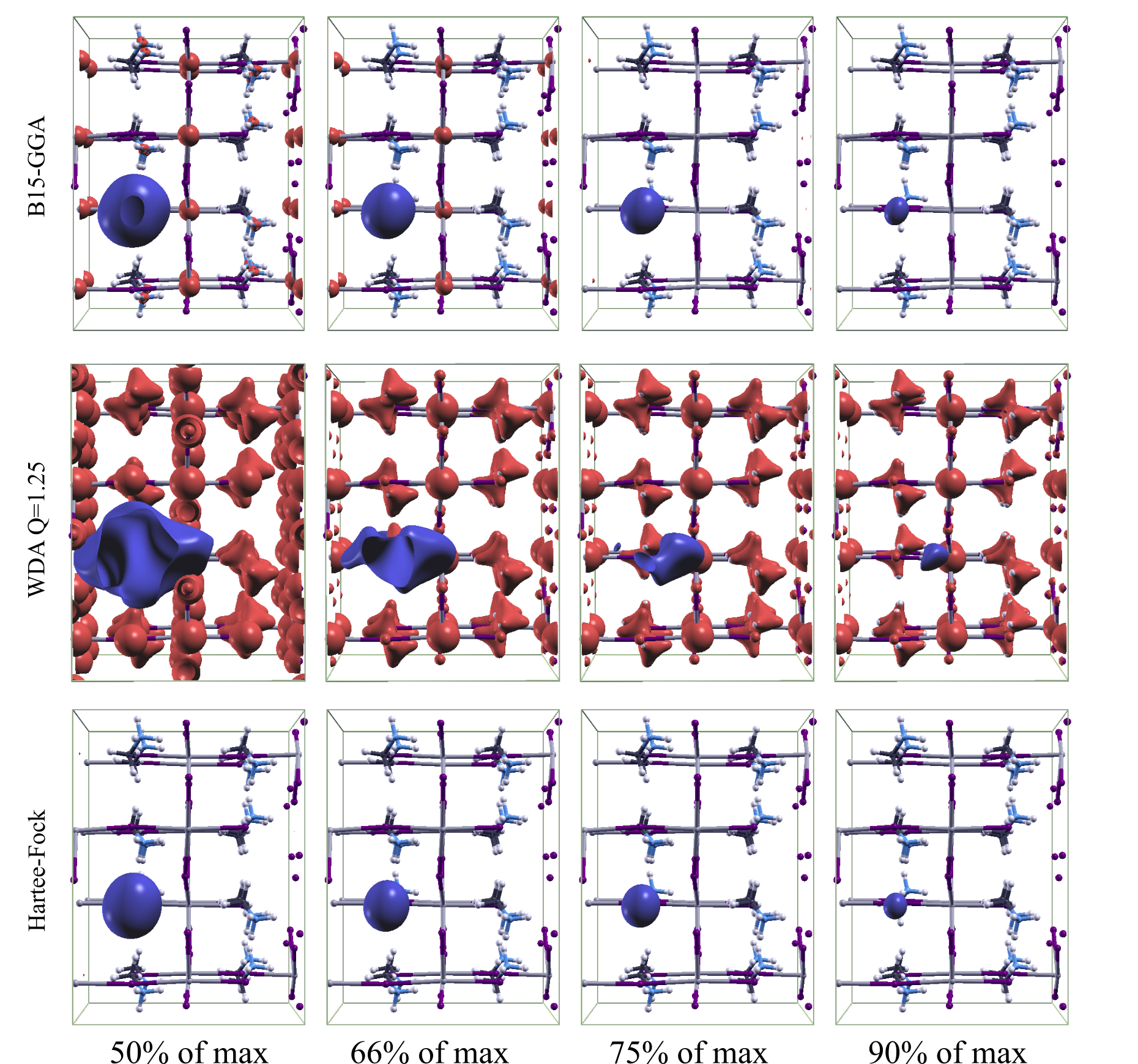}
	\caption{Comparison of positron density isosurfaces (blue) for a methylammonium vacancy in a 16 formula units tetragonal supercell with WDA (Q=1.25), B15-GGA and Hartree-Fock; here all the positron densities were scaled by a factor of 10. In this way the electron density (red) is visible only for the WDA. The maximum values of positron densities (before scaling) are -0.0174 for WDA, -0.105 for B15-GGA and -0.174 for HF (electrons/bohr$^3$). The maximum value of the electron density is 0.92 el/bohr$^3$.}
	\label{FigurePosDensities}
\end{figure}

We show here isosurfaces of the positron density for various enhancement factors in an alternative way as in the main paper. Instead of choosing, for each approximation, the same absolute value for the isosurface, we show them at the same percentage of the maximum value. In figure \ref{FigurePosDensities} all positron densities are multiplied by 10; the electron density is also shown, without any scaling factor. The Hartree-Fock result, i.e., with no electron-positron correlation, is shown for comparison.

Positron states corresponding to trapping in a vacancy should be localized, which correspond to a flat band in k-space. To check the localisation of our positron states in vacancies we calculated the positron eigenvalue at two {\bf k}-points: $\Gamma$ and L. The results are shown in table \ref{Table_Dispersion}, which shows that the dispersion of positron eigenvalues in both vacancies is fairly small, with higher localisation in methylammonium vacancies for semilocal enhancement factors. An exception occurs for the V$_{MA}$ with the WDA approximation, for which the dispersion is very large, and much larger than the lead vacancy.

	\begin{table}[h]
		\caption{Differences between positron eigenvalues at L and $\Gamma$ ($\epsilon_L-\epsilon_\Gamma$) in eV for various supercells with and without defects. All in tetragonal MAPbI$_3$ except for pristine polymorphous cubic (last line).}
		\begin{tabular}{lccc}\hline\hline
			defect/supercell & B15-GGA & B95-GGA      & WDA (Q=1.25) \\ \hline
			pristine (16 fu) & 0.393   & 0.389        & 0.419 \\       
			V$_{MA}^-$ (16 fu) & 0.030 & 0.028        & 0.227  \\       
		    V$_{Pb}^{--}$ (16 fu) & 0.066 & 0.058        & 0.063 \\ \hline
			pristine (32 fu) & 0.257   & ---          & ---     \\    
			V$_{MA}^-$ (32 fu) & 0.006 & ---          & ---      \\   
		    V$_{Pb}^{--}$ (32 fu) & 0.028 & ---          & ---      \\ \hline
		    pristine polymorphous (32 fu) & 0.259 & ---          & ---      \\ \hline\hline
		\end{tabular}
		\label{Table_Dispersion}
	\end{table}

\begin{figure}[h]
\includegraphics[width=0.9\columnwidth]{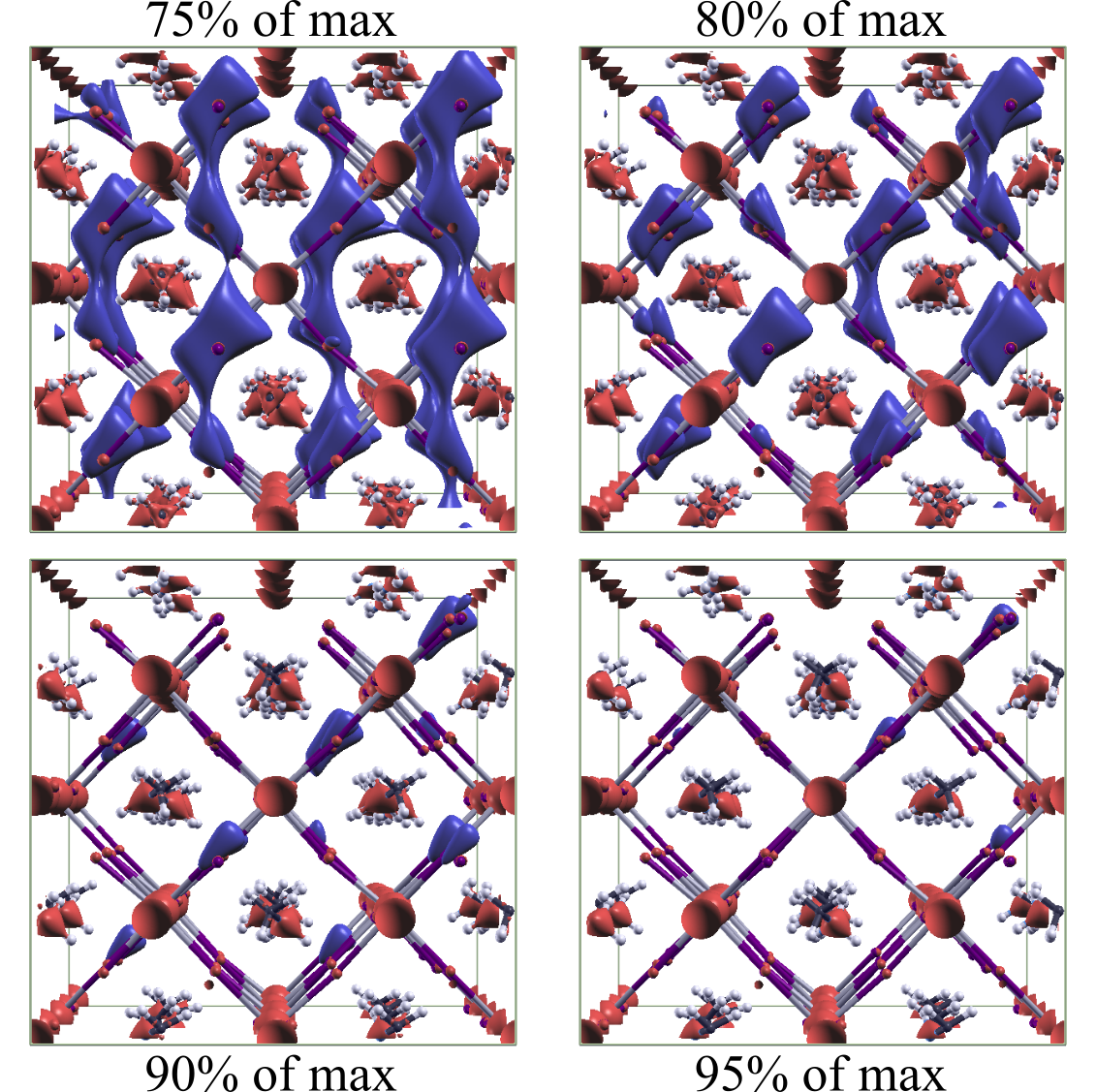}
	\caption{Positron densities in pristine (i.e., no defects) polymorphous MAPbI$_3$, as calculated with B15-GGA, with isosurfaces at 0.75, 0.8, 0.9, 0.95 of the maximum value. Positron densities (blue) are scaled by a factor of 50 to be comparable to the electron ones (red). }
	\label{FigurePolyDensities}
\end{figure}

To check whether positrons are able to localise in the polymorphous MAPbI$_3$ lattice or not (without vacancies), we checked the positron density for a polymorphous cubic cell. The results for isosurfaces at various fractions of the maximum values are shown in figure~\ref{FigurePolyDensities}. The influence of disorder on positron localisation seems to be a minor effect. 

\subsection{Determination of the average positron potential shift due to a defect}
\label{DeltaVpos}

\begin{figure}[h]
\includegraphics[width=0.9\columnwidth]{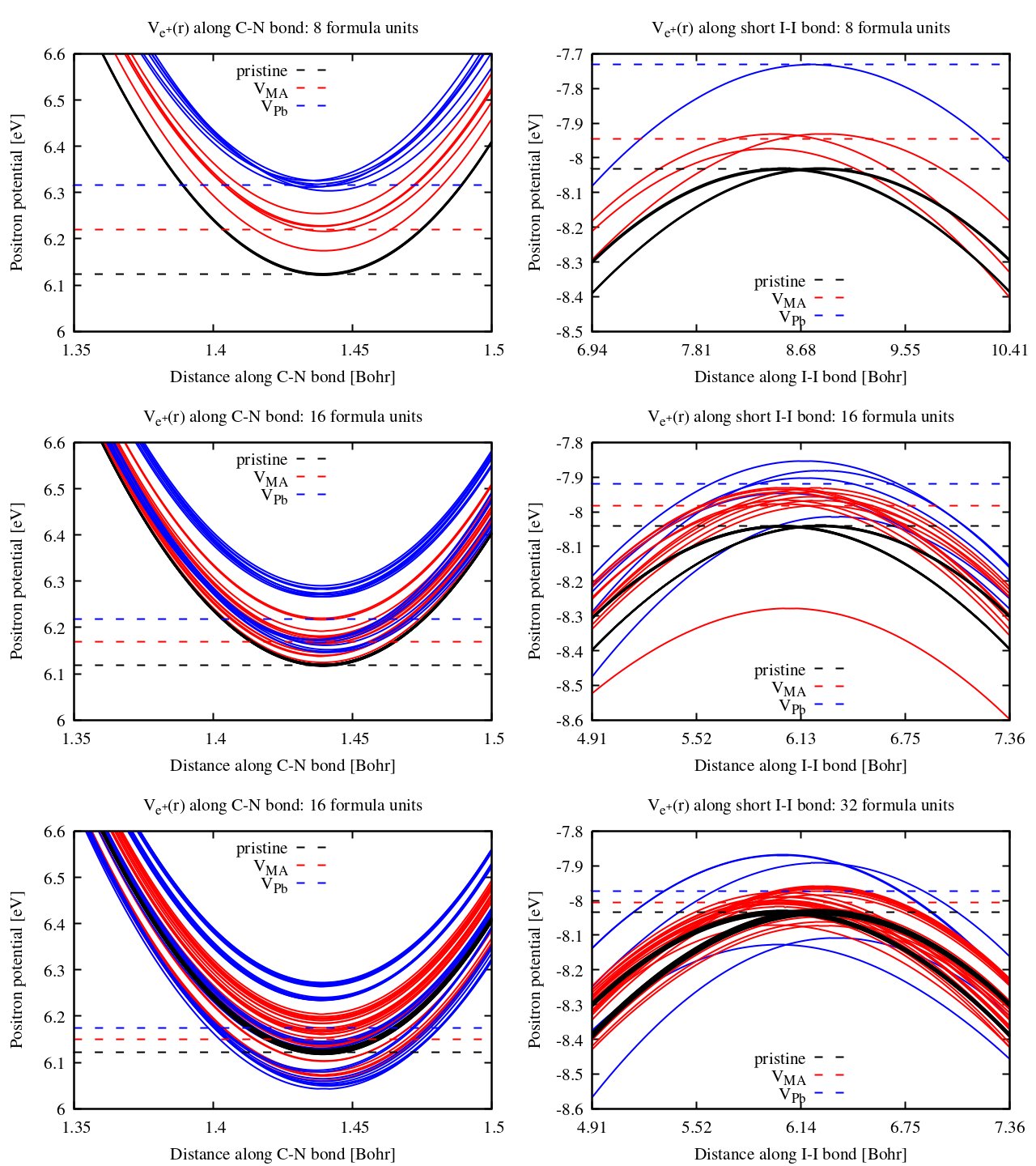}
\caption{Positron potential along C-N bonds (the three plots on the left) and in an interstitial region between two iodine atoms at distance 2.82 bohr (the three plots on the right side).
  Three supercell sizes are represented, as mentioned on the title of each plot. The dashed lines represent the average value of the minimum of the curve (for the left plots) or the of potential value at midpoint between the two iodines (right hand side plots). } 
	\label{FigureDV}
\end{figure}

The zero of the positron potential is arbitrary in our calculations. Thus, to compare eigenvalues from two different calculation we have to align the average positron potential. To do this we inspect the positron potential along C-N bonds of the methylammonium molecular ion, or in the midpoint between two iodine atoms whose distance is 12.27 bohr, and we take the average value.  
The results are shown in figure~\ref{FigureDV}. The values of the potential shifts obtained from the average of the minimum values along the C-N bonds is shown in table~\ref{TabDV}, for various enhancement factors and supercell sizes.

\begin{table}
  \caption{Positron potential shift for vacancies as calculated from the average of the minimum value along C-N bonds in supercells of three different sizes for tetragonal MAPbI$_3$ and various flavours of enhancement factors . Units are eV.}
  \begin{tabular}{llcccccccc}\\ \hline
     defect & cell size & B15-GGA & K14-GGA & B95-GGA & BN-LDA & WDA, Q=0.8 & WDA, Q=1.25 & WDA, Q=1.6\\ \hline 
V$_{MA}^-$ & 8fu  & 0.094  & 0.094  & 0.094  & 0.095  & 0.095  & - & 0.095 \\ 
V$_{MA}^-$ & 16fu  & 0.050  & 0.050  & 0.050  & 0.051  & 0.051  & 0.050 & - \\ 
V$_{MA}^-$ & 32fu  & 0.027  & 0.027  & 0.027  & 0.027  & -  & - & - \\ \hline 
V$_{Pb}^{--}$ & 8fu  & 0.193  & 0.192  & 0.193  & 0.196  & 0.194  & - & 0.189 \\ 
V$_{Pb}^{--}$ & 16fu  & 0.099  & 0.099  & 0.099  & 0.100  & 0.100  & 0.099 & - \\ 
V$_{Pb}^{--}$ & 32fu  & 0.051  & 0.051  & 0.051  & 0.052  & -  & - & - \\ \hline 
  \end{tabular}
  \label{TabDV}
  \end{table}

\subsection{Behavior of the positron potential on vacancy sites}

\begin{figure}[h]
\includegraphics[width=0.9\columnwidth]{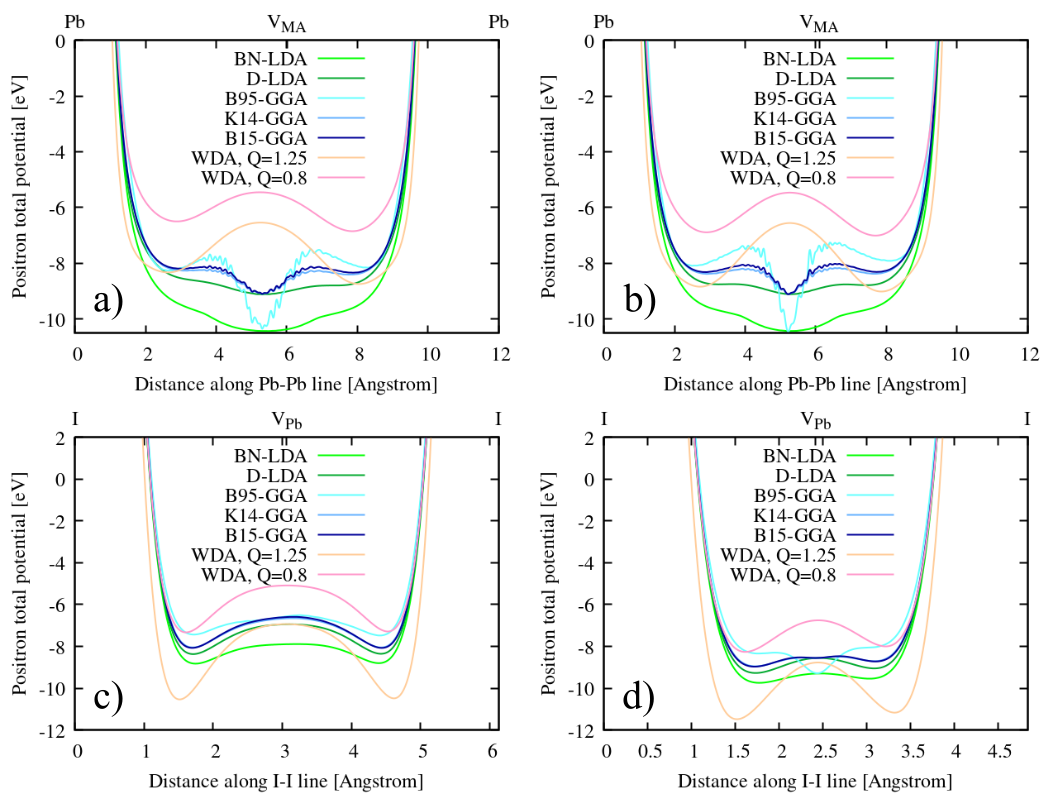}
\caption{Positron potential along a line passing through vacancy defects for several electron-positron correlation potentials. a) Positron potential along a line connecting two lead atoms adjacent to a methylammonium vacancy; b) same as a) with a different pair of lead atoms surrounding the same vacancy. c) Positron potential along a line connecting two iodine atoms adjacent to a lead vacancy; d) same as c) with a different pair of iodine atoms surrounding the same vacancy. } 
	\label{FigureVposVacancy}
\end{figure}

The behavior of the total positron potential at a vacancy site clearly distinguishes different approximations for the electron-positron correlation potential. In Fig.~\ref{FigureVposVacancy} we show the profile of the calculated positron potential along a line connecting two lead atoms adjacent to a methylammonium vacancy (panel a), and along a line between two of the six iodine atoms surrounding a lead vacancy. While all LDA and GGA approximations present a minimum on the vacancy site, the WDA approximation (here we show it for Q=0.8 and Q=1.25) goes through a slight maximum at the vacancy site.
This qualitative difference, together with the quantitative difference between the two vacancies, might explain the fact that the binding of the positron to V$_{MA}$ is stronger than to V$_{Pb}$ for the semilocal functionals, while the reverse holds for the non local WDA approximation, whatever the chosen value of the screening charge.

\FloatBarrier
\bibliography{PASpero}